\documentclass{aa}  

\usepackage[normalem]{ulem}
\usepackage{graphicx}
\usepackage{color}
\usepackage{url}
\usepackage{hyperref}
\usepackage{lipsum}
\usepackage{multirow}
\usepackage{xcolor}

\usepackage{txfonts}
\usepackage{natbib}
\usepackage{amsmath}    
\usepackage{amssymb} 
\usepackage{bm}
\usepackage{mathtools}

\usepackage{orcidlink}
\usepackage{xspace}

\usepackage[switch]{linenoaa} 
\newcommand{\source}{\mbox{Swift~J1727.8$-$1613}\xspace}

\begin{document} 

\title{Optical polarimetry of the accreting black hole X-ray binary Swift~J1727.8$-$1613 over the state transition and radio ejections}

\titlerunning{Optical polarimetry of Swift~J1727.8$-$1613}

\authorrunning{Nitindala, A.~P., et al.}

\author{Anagha P. Nitindala\inst{1}\orcidlink{0009-0002-7109-0202}
\and
Alexandra Veledina\inst{1,2}\orcidlink{0000-0002-5767-7253} 
\and Vadim Kravtsov\inst{1}\orcidlink{0000-0002-7502-3173}
\and Andrei V. Berdyugin\inst{1}\orcidlink{0000-0002-9353-5164}
\and Mar\'ia~Alejandra~D\'iaz~Teodori\inst{1}\orcidlink{0009-0002-1852-7671}
\and Vilppu Piirola\inst{1}\orcidlink{0000-0003-0186-206X}
\and Takeshi Sakanoi\inst{3}\orcidlink{0000-0002-7146-9020} 
\and Masato~Kagitani\inst{3}
\and Svetlana~V.~Berdyugina\inst{4}\orcidlink{0000-0002-2238-7416}
\and Juri~Poutanen\inst{1}\orcidlink{0000-0002-0983-0049}}

\institute{Department of Physics and Astronomy, FI-20014 University of Turku, Finland\\
\email{anaghapradeep.a.nitindala@utu.fi}
\and Nordita, Stockholm University and KTH Royal Institute of Technology, Hannes Alfv\'ens v\"ag 12, SE-10691 Stockholm, Sweden 
\and Graduate School of Sciences, Tohoku University, Aoba-ku, 980-8578 Sendai, Japan 
\and Istituto ricerche solari Aldo e Cele Dacc\'o (IRSOL), Faculty of Informatics, Universit\`a della Svizzera italiana, Via Patocchi 57, Locarno, Switzerland
}

\abstract
{We present the first optical ($BVR$) polarimetric observations of \source during its 2023--2024 outburst. 
Observations were performed during the X-ray hard-to-soft state transition, the soft state, and the decaying hard state of the source.
For the vast majority of nights, we detect statistically significant polarization of ${\approx}1$\%, a fraction of which is of interstellar origin.
We find a significant change of polarization coinciding in time with discrete radio ejections.
The direction of this polarization variation differs from the directions inferred from the X-ray, submillimeter, and radio polarization angles, as well as from the resolved jet orientation.
After correcting for the interstellar component, we find that the intrinsic polarization degree remained approximately constant at PD $\approx 0.3$\% throughout the hard-intermediate state.
We explore several possible origins for the polarization and conclude that it is most plausibly produced by scattering within the optically thin accretion disk wind.
The intrinsic polarization angle, PA~$\approx-15\degr$, is notably offset from the jet axis, which we interpret as evidence of a misalignment between the black hole spin and the orbital axis.}

\keywords{accretion, accretion disks -- polarization -- stars: black holes -- individuals: \source}
\maketitle


\section{Introduction}

During outbursts, black hole (BH) X-ray binaries undergo transitions between several distinct spectral states.
At the rising phase of an outburst, the sources are typically found in the hard spectral state, where the X-ray spectrum is dominated by Comptonization in a hot, optically thin plasma, often referred to as the hot accretion flow or corona \citep{ST80,PS96,PKR97,Esin97}.
Thermal emission from the (irradiated) optically thick accretion disk \citep{SS73, NT73, PT1974} dominates the ultraviolet to optical wavelengths. 
In addition, synchrotron emission from nonthermal particles, coming either from the hot accretion flow itself or from the relativistic jet, contributes to the optical and infrared (OIR) bands \citep{Poutanen2014,UttleyCasella2014}.
Superimposed on the continuum, hard-state BHs often exhibit emission or absorption lines in their UV and OIR spectra, which are typically associated with the disk or its winds \citep{Munoz-Darias2016,Sanchez-Sierras2020,CastroSegura2022}.
Bright radio emission is also observed, produced by the compact jet.

During the transition to the soft state (SS), through the so-called hard- and soft-intermediate states \citep[HIMS and SIMS;][]{Homan2005}, the contribution of the hot medium to the X-ray spectrum gradually diminishes.
The nonthermal synchrotron component in the OIR range likewise decreases substantially, while the optically thick disk emission becomes dominant across the broadband OIR--X-ray spectrum.
The jet is quenched following a brief phase of enhanced radio emission associated with discrete ejection events.
Wind signatures have been detected in the infrared, but not in the optical, where visibility is likely affected by ionization \citep{MataSanchez2018,Sanchez-Sierras2020}.
During the reverse transition, the contributions of the hot accretion flow and/or jet emission increase again, becoming dominant in the broadband spectral energy distribution over the entire decaying hard state (DHS).

Although the general sequence of spectral and timing changes during outbursts is well established, several key questions remain unresolved.
These include the relative contributions of the accretion flow and jet to the OIR emission, the role of winds in angular momentum and mass transfer, and the relative orientation between the outer disk, the inner accretion flow, and the jet.
Measurements of optical polarization and its evolution throughout the outburst, along with spectral and timing information, offer valuable diagnostics to distinguish between different emission mechanisms, thereby constraining the nature of the accretion engine and its feedback on the interstellar medium.

\begin{figure*}
\centering
\includegraphics[width=0.9\linewidth]{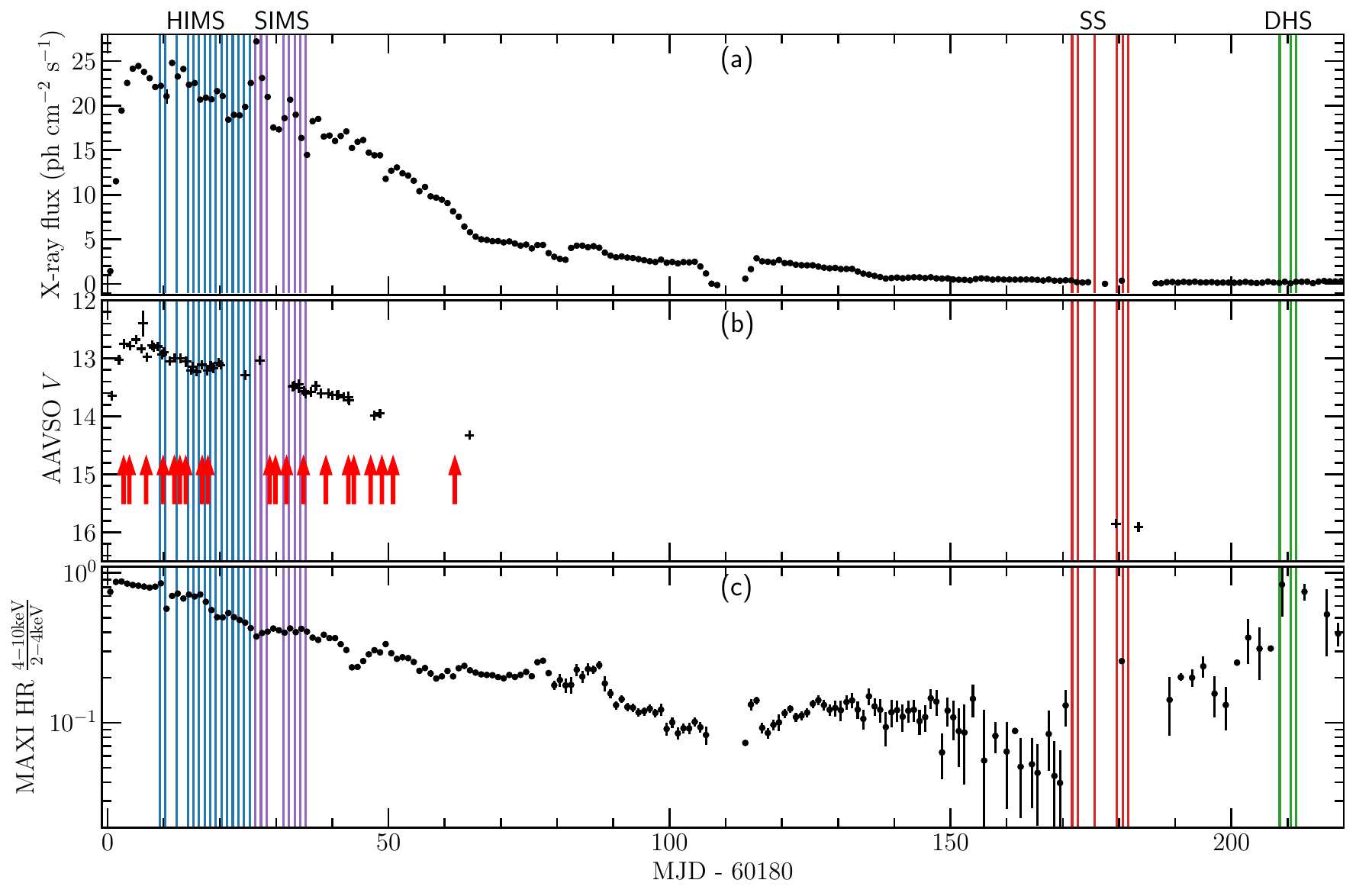}
\caption{X-ray and optical light curves and X-ray HR of \source. {\it Panel (a)}: MAXI 2--20 keV X-ray light curve. {\it Panel~(b)}: AAVSO Johnson $V$ filter light curve (\url{https://www.aavso.org}). The red arrows indicate the dates of optical spectroscopic detections of the accretion disk winds \citep{MataSanchez2024outflows}. {\it Panel (c)}: MAXI 4--10 keV/2--4 keV HR. 
Solid blue, purple, red, and green lines indicate dates of optical polarimetric observations through the hard intermediate-, soft intermediate-, soft-, and decaying hard-states of the outburst, respectively.}
\label{fig:lc}%
\end{figure*}

The polarization degree (PD) depends on both the emission mechanism and the system inclination.
For the optically thick accretion disk, polarization arises from electron scattering in the disk atmosphere and can reach up to 11.7\% for edge-on systems \citep{Cha60,Sob63,Rees1975}, but rapidly decreases to zero for the face-on systems.
In contrast, synchrotron radiation can exhibit much higher polarization levels, reaching ${\sim}$70--75\% in ideal cases \citep{GinzburgSyrovatskii1965}, but the PD strongly depends on the degree of magnetic field ordering, becoming zero for a tangled field topology.
Furthermore, polarization may arise from scattering in the optically thin medium \citep{ST85}, either in the accretion disk wind \citep{Nitindala2025}, or by Comptonization processes in the optically thin accretion flow \citep{PS96,Poutanen2023}.
The latter case gives rise to X-ray polarization that can serve a fine diagnostic of emission processes and accretion geometry in the immediate BH vicinity \citep[e.g.,][]{Krawczynski22,Veledina2023,Ewing2025}.

The polarization angle (PA) provides additional information on the geometry of the emission region.
Synchrotron emission is polarized in the direction orthogonal to the projection of the magnetic field lines on the sky \citep{GinzburgSyrovatskii1965}.
For the scattering events in the optically thin or thick cases, the PA is either aligned with or orthogonal to the axis of the system \citep{ST85}, such as that of the accretion disk, the jet, or the BH spin.
Deviations from these axes may indicate the importance of relativistic aberration or gravitational lensing effects \citep{StarkConnors1977,ConnorsPiranStark1980}, or a nonzero angle between the aforementioned directions \citep{Poutanen2022}.

Optical polarimetric monitoring of several BH X-ray binaries during the outburst has revealed the presence of intrinsic polarization components \citep[e.g.,][]{Boyd2001,Schultz2004,Tanaka2016,Kosenkov2017,Kravtsov2019ATel,Kravtsov2023CygX1,Kravtsov2022,Mastroserio2025,Rout2025}.
A number of sources were likewise studied in optical polarized light in quiescence \citep[e.g.,][]{Dubus2008,Kravtsov2022,Kravtsov2025}. 
Perhaps the most comprehensive optical polarimetric coverage of the outburst was obtained for the unusually bright, high-inclination \citep{Torres2020,Wood2021} BH X-ray binary \mbox{MAXI J1820+070}.
A sub-percent level of intrinsic polarization, oriented parallel to the jet axis during the rising hard state, disappeared in the soft and decaying hard states, coinciding with the emergence or disappearance of wind signatures \citep{Veledina2019,Kosenkov2020}.
The subsequent discovery of the PA changes during the approach of quiescence suggested a considerable misalignment between the BH spin and orbital axis in this system \citep{Poutanen2022}.
In this paper, we study the evolution of optical polarization signatures in a recent bright X-ray transient \source.

\source is a low-mass X-ray binary discovered in August 2023 \citep{Page2023J1727detection,Negoro2023detection}. 
After a brief rising hard state (RHS) phase of less than a week, the system spent over two months in the HIMS--SIMS (see Fig.~\ref{fig:lc}), exhibiting several flaring episodes accompanied by discrete radio ejections \citep{Wood2025jets}.
Optical spectroscopy taken in subsequent quiescence revealed a BH primary in a binary with a K-type companion, and an orbital period of $10.804\pm0.001$~h \citep{MataSanchez2025}. 
Optical signatures of disk winds were detected across the outburst \citep{MataSanchez2024outflows}. 
Their properties, along with the absence of detectable X-ray dips, suggest a low-to-moderate orbital inclination, $\lesssim60\degr$, of the system \citep{MataSanchez2025}.
The observer inclination relative to the jet was likewise constrained to be moderate, $<74\degr$ \citep{Wood2024}.

\begin{table*}
\caption{Weighted average observed optical polarization of \source in the different spectral states.}             
\label{table:avgobs}      
\centering          
\begin{tabular}{lccccccc}     
\hline\hline       
&  & \multicolumn{2}{c}{$B$} & \multicolumn{2}{c}{$V$} & \multicolumn{2}{c}{$R$}\\ 
State & MJD & PD (\%) & PA (deg) & PD (\%) & PA (deg) & PD (\%) & PA (deg) \\
\hline
HIMS & 60189--60205 & 0.917 $\pm$ 0.001  & 72.11 $\pm$ 0.03 & 0.929 $\pm$ 0.002 & 68.73 $\pm$ 0.06  & 0.797 $\pm$ 0.001 & 66.46 $\pm$ 0.04 \\
SIMS & 60206--60215 & 1.104 $\pm$ 0.001 & 72.09 $\pm$ 0.03 & 1.122 $\pm$ 0.007 & 68.2 $\pm$ 0.2 & 1.107 $\pm$ 0.003 & 69.76 $\pm$ 0.07 \\
HIMS+SIMS & 60189--60215 & 0.947 $\pm$ 0.001  & 72.11 $\pm$ 0.03 & 0.962 $\pm$ 0.002 & 68.63 $\pm$ 0.07 & 0.871 $\pm$ 0.002 & 67.47 $\pm$ 0.06 \\
\hline
SS & 60351--60361 & 1.18 $\pm$ 0.11 & 76.5 $\pm$ 2.7 & 1.14 $\pm$ 0.10 & 63.0 $\pm$ 2.5 & 1.01 $\pm$ 0.05 & 59.9 $\pm$ 1.4 \\
\hline
DHS & 60388--60391 & 1.05 $\pm$ 0.07 & 71.1 $\pm$ 2.0 & 1.67 $\pm$ 0.09 & 69.4 $\pm$ 1.5  & 1.10 $\pm$ 0.08 & 70.6 $\pm$ 2.0 \\
\hline                
\end{tabular} 
\tablefoot{The full list of detections is given in Table~\ref{table:obs}.}
\end{table*}

X-ray polarization was first measured at the beginning of transition toward the SS and was found to be PD${=}4.1\pm0.2$\%  at PA${=}2\fdg2\pm1\fdg3$ \citep{Veledina2023}.
The X-ray PA was found to be consistent with those measured in the radio, sub-millimeter, and optical bands \citep{Vrtilek2023atel,Kravtsov2023atel}, and was also aligned with the position angle of the jet axis \citep{Wood2024}.
The X-ray PD decreased over the HIMS--SIMS, becoming ${\lesssim}$1\% as the source entered the SS, and recovered to $3.3\pm0.4$\% in the DHS \citep{Ingram2024,Svoboda2024swift,Podgorny2024}. 
The relatively high PD observed in the hard states, for a source at low-to-moderate inclination, and similarity of the polarization properties for the rise and decay, despite the luminosity difference of two orders of magnitude, suggest that scattering of central radiation in accretion disk winds may play a significant role in producing the observed polarization \citep{Nitindala2025}.
The wind may also give a substantial contribution to the optical polarization, giving the correlated appearance of X-ray and optical polarimetric properties, along with the wind signatures in the broadband spectra.

In this work, we present the first study of the evolution of optical polarization throughout the outburst of the binary, covering the HIMS, SIMS, SS, and DHS.
The data used in the paper are described in Sect.~\ref{sec:data}. 
The results of our analysis is presented in Sect.~\ref{sec:results} and their interpretation is discussed in Sect.~\ref{sec:discuss}. 
Finally, we summarize our findings in Sect.~\ref{sec:summary}.

\section{Observations and data reduction}
\label{sec:data}

Optical polarimetric observations of \source were performed with the high-precision DIPol-2 polarimeter \citep{Piirola2014Dipol2} mounted on the remotely controlled Tohuku~60\,cm (T60) telescope at the Haleakala Observatory, Hawaii. 
DIPol-2 is a double-image polarimeter equipped with a rotatable super-achromatic half-wave plate as a polarization modulator and plane-parallel calcite as a polarization two-beam analyzer. 
Two dichroic sharp-pass filters split (doubled) light beam into three passbands.  
This allows one to measure polarization in three $BVR$ bands simultaneously with three individual CCD cameras. 
With this design, orthogonally polarized sky images overlap, and the sky polarization is completely eliminated, even if it is variable. 
Accuracy of DIPol-2 is only photon-limited and polarization measurement with a precision of up to $10^{-5}$ is routinely reached for sufficiently bright stars. 

We observed \source for a total of 33 nights between August 2023 and April 2024, with 80 to 200 individual measurements of normalized Stokes $q$ and $u$ parameters per night.
For each night, we computed the average Stokes parameters using the $2\sigma$ iterative weighting algorithm \citep{Kosenkov2017,Piirola2020}
and the resulting PD, 
\begin{equation} 
\label{eq:PD}
P=\sqrt{q^2 + u^2} 
,\end{equation}
and PA as 
\begin{equation} 
\label{eq:PA}
\theta=\frac{1}{2}{\rm atan2}(u,q) , 
\end{equation}
with atan2 being the quadrant-preserving arc tangent.  
The error on the PD ($\sigma_P$) is the uncertainty on the individual Stokes parameters and the error on the PA for significant detections was estimated as \citep[e.g.,][]{Clarke1986,Naghizadeh1993} 
\begin{equation} 
\label{eq:errPA}
\sigma_{\theta} = \frac{\sigma_P}{2P} . 
\end{equation}
To calibrate the instrumental polarization, we observed 12 nearby unpolarized stars from the list given in \citet{Piirola2020}. 
The instrumental polarization was found to be on the order of $\approx 0.001\%$ in all pass-bands, and hence was considered to be negligible. 
The zero point of the PA was determined by observing highly polarized standard stars HD~161056 and HD~204827. 
Observations of the field stars were carried out to reliably determine the interstellar (IS) polarization using both DIPol-UF mounted on the Nordic Optical Telescope, La Palma, and DIPol-2 polarimeters. Similar to DIPol-2, instrumental polarization was calibrated for DIPol-UF as well, using six nearby unpolarized stars listed in \citet{Piirola2020} and the same highly polarized stars.

\begin{figure}
\centering
\includegraphics[width=0.85\linewidth]{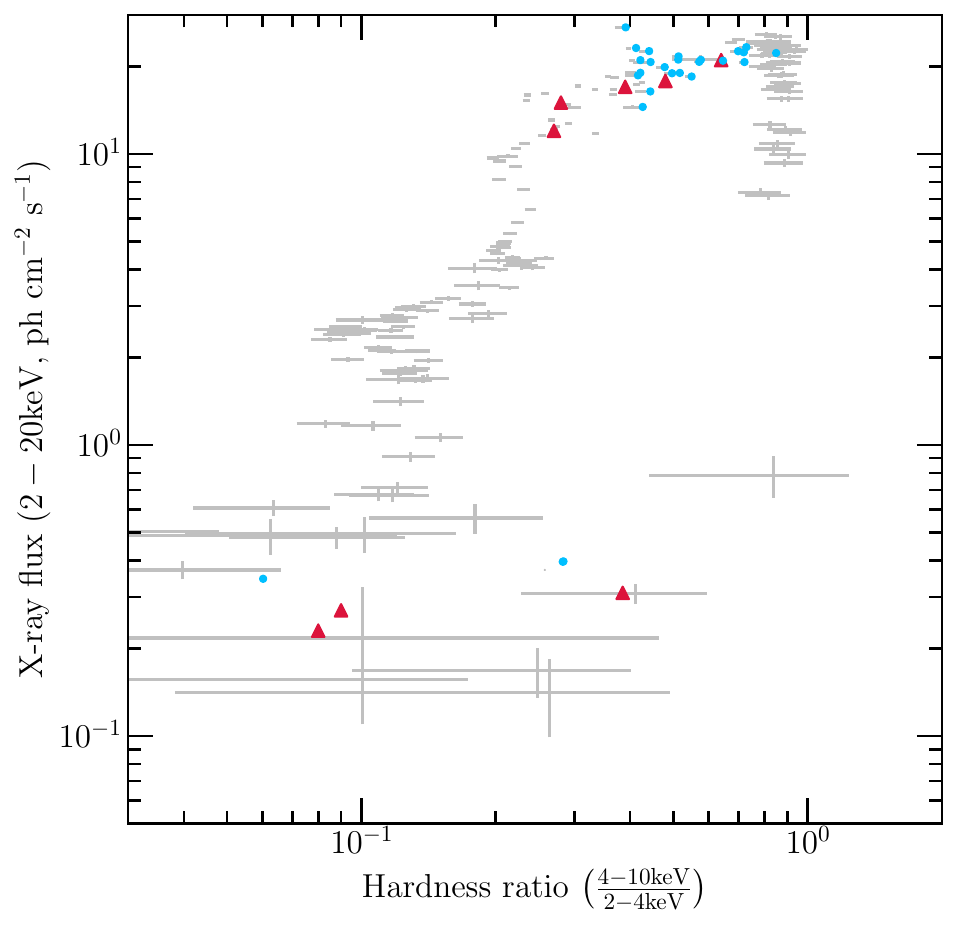}
\caption{Hardness-intensity diagram of \source. 
The gray crosses are from MAXI data. The hardness is taken as the ratio of photon fluxes between 4--10 and 2--4 keV bands. 
The X-ray photon flux is taken in the 2--20 keV range. 
The blue circles correspond to the times of optical polarimetric observations. 
For some optical measurements, the corresponding HR could not be determined, and hence are not shown in the figure. 
The red triangles denote position of the source during X-ray polarimetric observations by IXPE \citep{Podgorny2024}.}
\label{fig:hid}
\end{figure}

Optical photometric data were obtained from the AAVSO\footnote{\url{https://www.aavso.org/}} database in the Johnson $V$ band, binned to 1-day intervals. 
The resulting light curve is shown in Fig.~\ref{fig:lc}b.
Additional photometric measurements in the $BVR$ bands were extracted from our polarimetric observations. 
Two field stars seen in the images were used to perform differential photometry. 
Their magnitudes in the respective bands were calculated using {\it Gaia}\footnote{\url{https://www.cosmos.esa.int/web/gaia}} DR3 data. 
While these measurements might suffer from uncertainties due to possible variations in fluxes of the field stars and/or calibration uncertainty of {\it Gaia} $G$ magnitudes to $BVR$ magnitudes (especially in the $B$ band), the $V$-band data points were found to be in good agreement with the AAVSO data.

The X-ray flux was obtained using the Monitor of the All-sky X-ray Image \citep[MAXI;][]{Matsuoka2009} in the 2--4, 4--10, 10--20, and 2--20 keV bands.\footnote{\url{http://maxi.riken.jp/star_data/J1727-162/J1727-162.html}} 
The light curve is shown in Fig.~\ref{fig:lc}a and the calculated hardness ratio (HR) is shown in Figs.~\ref{fig:lc}c and \ref{fig:hid}. 
The HR is used to differentiate between different spectral states of the source during its outburst phase (see Fig.~\ref{fig:hid}). 
X-ray polarimetric data through the different states obtained with IXPE  \citep{Veledina2023,Ingram2024,Podgorny2024,Svoboda2024swift} were also used to interpret our results.

\section{Results}
\label{sec:results}

\subsection{Observed polarization}\label{sec:obs_pol}

Figure~\ref{fig:lc} shows the X-ray flux, optical magnitude, and X-ray HR of \source through the outburst. Optical polarimetric observations (shown using vertical colored lines; see Table~\ref{table:obs}) have been grouped into different spectral states (HIMS, SIMS, SS, and DHS) based on the X-ray flux and the HR, as well as earlier reports of state transitions \citep{2023ATel16247Bollemeijer,2023ATel16271MillerJones,2024ATel16541Podgorny}.
Some of these observations were contemporaneous with spectroscopic wind detections (\citealt{MataSanchez2024outflows}; marked by red arrows in Fig.~\ref{fig:lc}b) and with X-ray polarimetric observations (see Fig.~\ref{fig:hid}).

\begin{figure}
\centering
\includegraphics[width=0.9\linewidth]{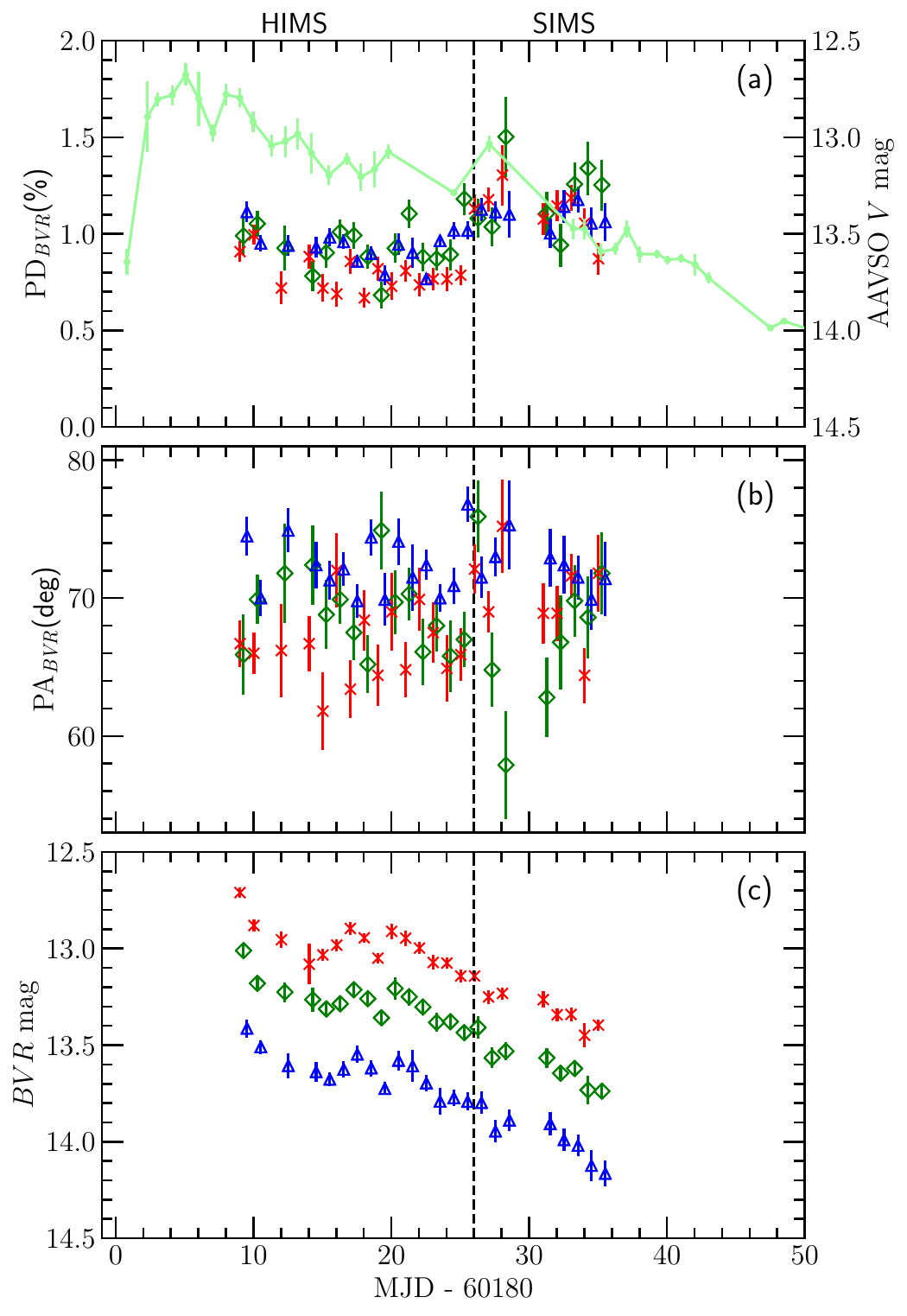}
\caption{Evolution of \source optical polarization and $BVR$ magnitudes around the HIMS/SIMS state transition. The colored symbols correspond to the $B$ (blue triangles), $V$ (green diamonds), and $R$ (red crosses) bands.
\textit{Panel (a)}: AAVSO light curve (light green line) and the observed optical PD in three filters. 
\textit{Panel (b)}: Observed PA.
\textit{Panel (c)}: Optical magnitudes calculated from DIPol-2 images as described in the text. 
The vertical dashed black line corresponds to MJD 60206, when the jet ejections were detected \citep{Wood2025jets}.}
\label{fig:lc_pol}
\end{figure}

\begin{figure}
\centering
\includegraphics[width=0.84\linewidth]{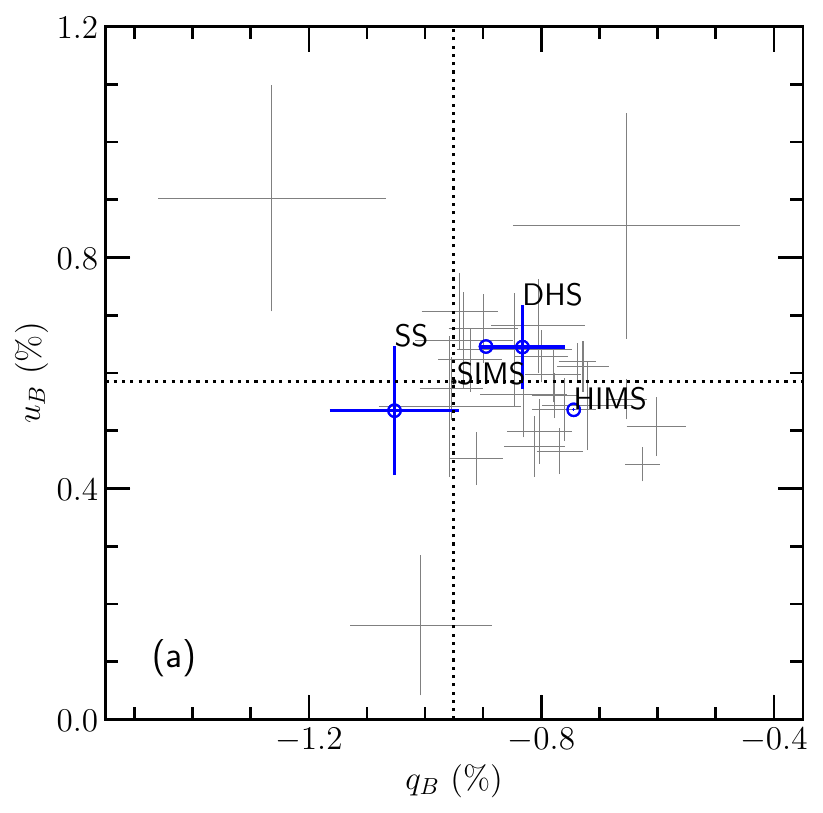}
\includegraphics[width=0.84\linewidth]{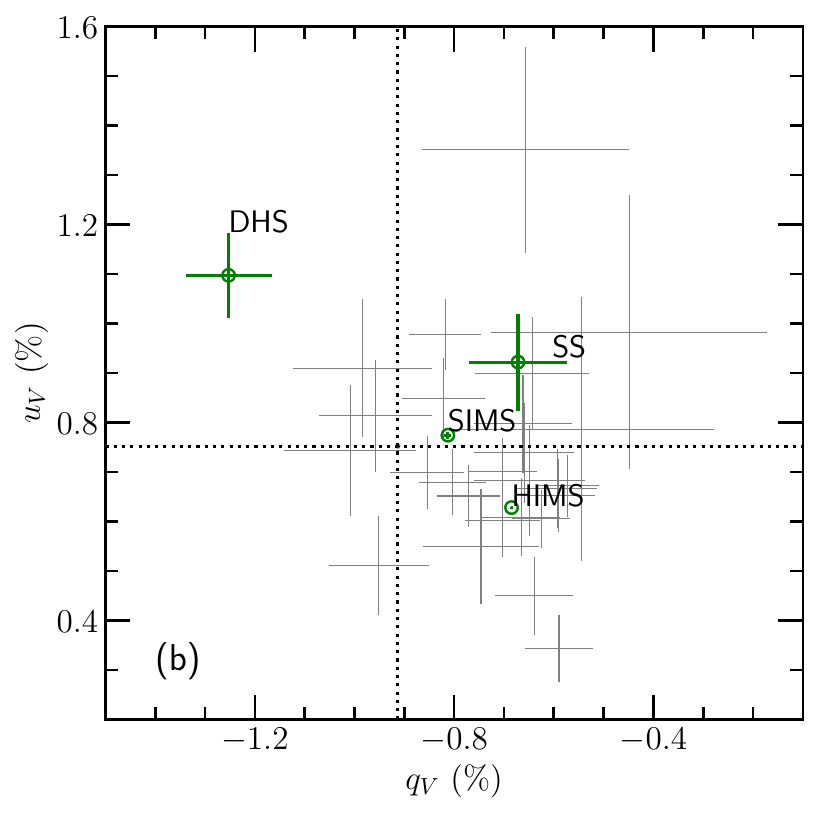}
\includegraphics[width=0.84\linewidth]{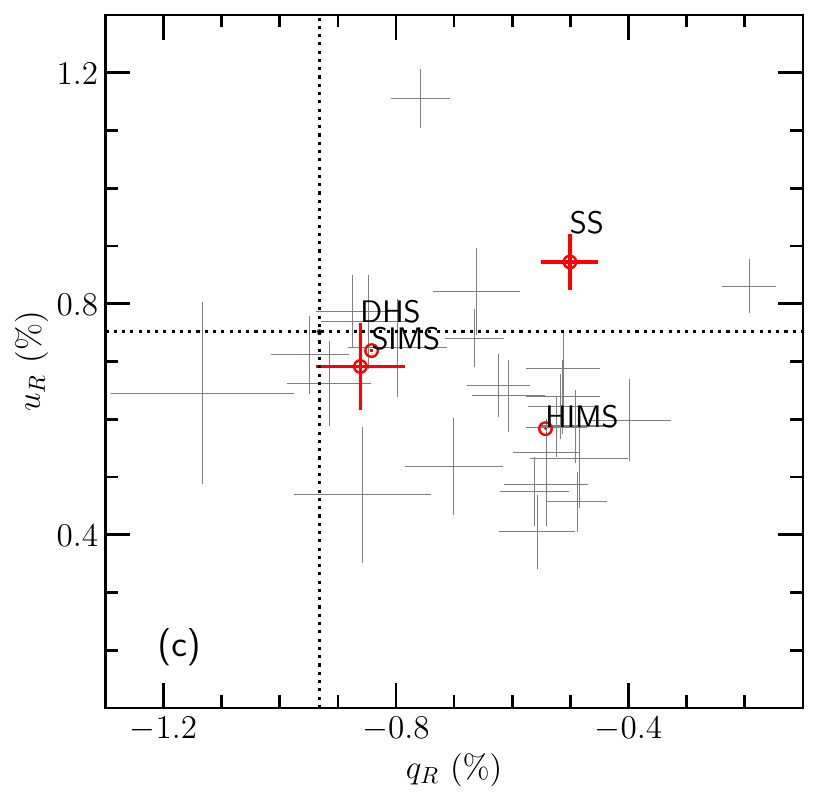}
\caption{Observed normalized Stokes parameters $q$ and $u$  of \source (gray crosses), along with their state-wise averages (colored, open symbols; as marked in the figure) in $B$ (panel a), $V$ (panel b), and $R$ (panel c) bands. 
The errors on observed data of the source are $1\sigma$. 
The IS polarization estimate is given by the black circle ($3\sigma$ error) at the intersection of the dotted lines.}
\label{fig:qu_all}
\end{figure}

The evolution of the observed optical polarization of the source during its transition toward the SS is shown in Fig.~\ref{fig:lc_pol}. 
We find that the PD increases after MJD 60206 (vertical dashed line) in all bands. 
This coincides with the date when radio ejections were detected \citep{Wood2025jets} in the source, which also corresponds to an X-ray flaring event (Fig.~\ref{fig:lc}a).
Thus, we further split the transition into HIMS and SIMS, which consist of observations before and after (including) MJD 60206.
We note that the incursion into SIMS, marked by the appearance of type-B quasi-periodic oscillations, was reported at the time of radio ejections \citep{Mereminskiy2024}. 
However, the source subsequently returned to the HIMS, and the exact date of the SIMS-HIMS transition is uncertain. Therefore, we refer to the observing period following the ejections as the SIMS.
The observed PA also behaves differently after this date compared to before, especially in the $V$ and $R$ bands. 
The weighted average polarization values during the two states identified above, as well as the entire transition, are given in Table~\ref{table:avgobs}. 
The weighted average during each state was computed by taking the weights to be inversely proportional to the square of the error on the nightly means that are given in Table~\ref{table:obs} and the corresponding error on the averages is the standard deviation around the weighted average.
The average Stokes parameters are given in Fig.~\ref{fig:qu_all}.
The evolution of polarization between HIMS and SIMS is apparent in all bands.

\begin{figure}
\centering
\includegraphics[width=0.9\linewidth]{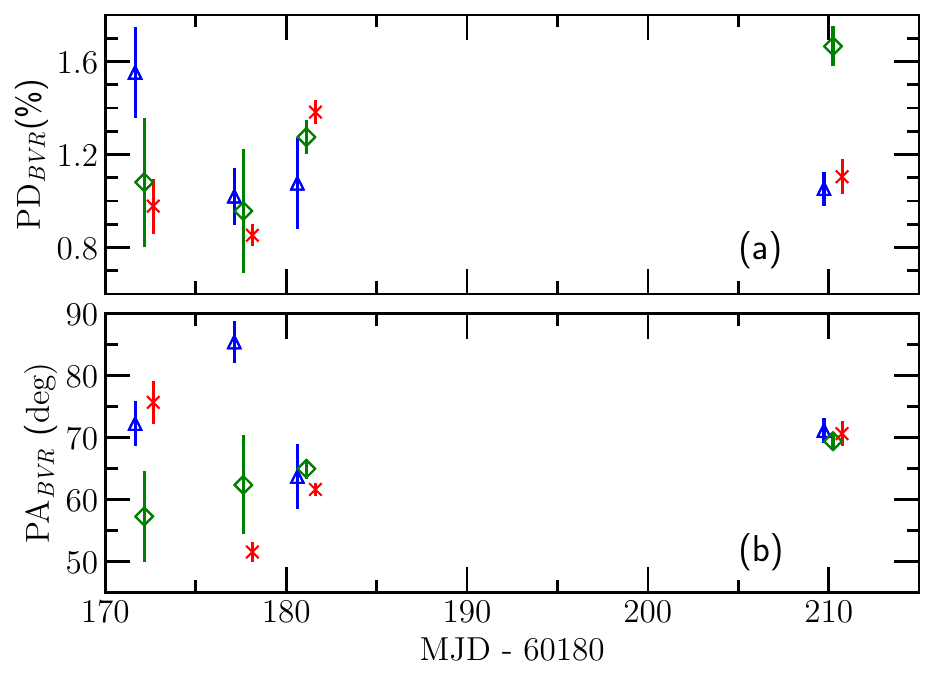}
\caption{Evolution of the observed PD (panel a) and PA (panel b) in SS and DHS is shown in blue triangles, green diamonds, and red crosses for the $B$, $V$, and $R$ bands, respectively.}
\label{fig:PDPA_SSDHS}
\end{figure}

To statistically quantify the significance of the polarization change, we used the multivariate Hotelling's $T^2$ test \citep{Hotelling1931} for two variables, $q$ and $u$.
The details of applying this test to optical polarimetric data have been described in \citet{Kosenkov2017}. 
In the $B$ band, we obtained a $t^2$ value of 21.3, giving an F statistic, $f= 10.6$ at a p value of $p < 2\times10^{-4}$.
The total number of observations, including both sets of data, is 918. 
This gives the probability that the polarization is equal before and after MJD 60206 to be $2.7 \times 10^{-5}$. 
Similarly, in the $V$ and $R$ bands, we obtained $t^2 = 7.7$ and 28.5, $f = 3.9$ and 14.2, and $p=0.0219$ and $<2\times10^{-5}$, resulting in probabilities that the polarization is equal before and after MJD 60206 to be 0.0215 and $8.2 \times 10^{-7}$, respectively.
Thus, we find a statistically significant change of polarization between HIMS and SIMS in the $B$ and $R$ bands. 

As the source transitioned into the SS and subsequently into the DHS, its brightness decreased substantially, resulting in larger statistical uncertainties in the nightly mean measurements (see Table~\ref{table:obs}). 
To improve the signal-to-noise ratio and obtain  significant detections at the $>3\sigma$ level, we combined data from consecutive nights, producing two-night mean polarization measurements for the SS and three-night means for the DHS.
The temporal evolution of these measurements is shown in  Fig.~\ref{fig:PDPA_SSDHS}, while their average values are summarized in Table~\ref{table:avgobs} and illustrated in Fig.~\ref{fig:qu_all}.
Due to the comparatively large uncertainties in these states relative to those measured during the HIMS--SIMS, no statistically significant differences in polarization between states can be established.

\begin{figure}
\centering
\includegraphics[width=0.9\linewidth]{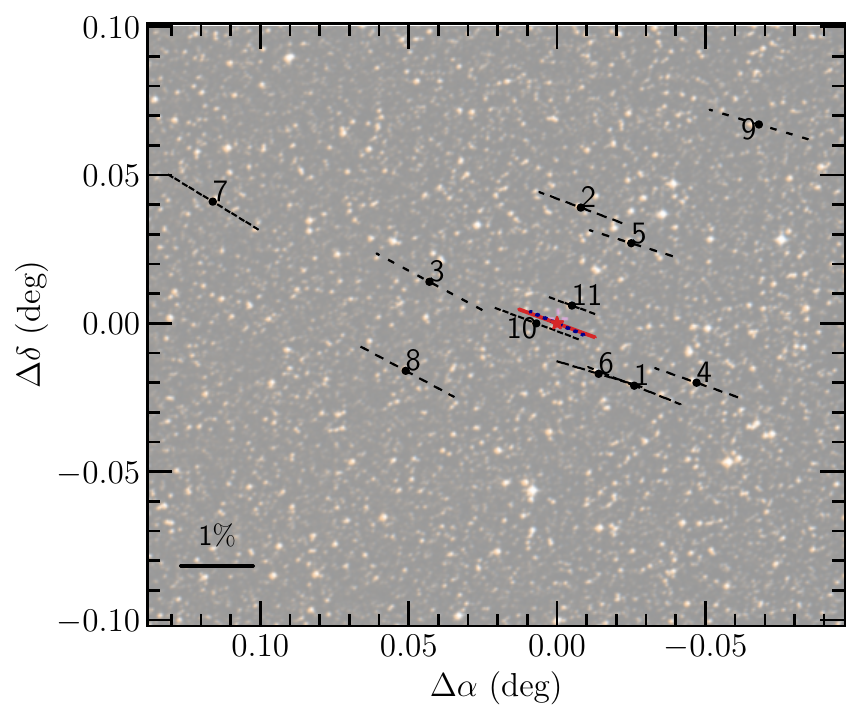}
\caption{Observed polarization of \source (red star) and the field stars (black circles) on the sky in the $R$ band. The length of the bar corresponds to the PD and is denoted by dashed black lines for the field stars. For \source, the solid red line and the dotted blue line correspond to the PD during SIMS and HIMS, respectively. The directions of the bars denote the PA. The background image is taken from the Aladin sky atlas (\url{https://aladin.cds.unistra.fr/}).}
\label{fig:sky}
\end{figure}

Thus, we detect significant polarization variability only during the HIMS-SIMS transition, following a flaring event.
This indicates the presence of polarization that is intrinsic to the source.
Assuming that the observed variability arises solely from changes in the PD, while the intrinsic PA remains constant, the direction of the shift from HIMS to SIMS on the $q$–$u$ plane provides an estimate of the intrinsic PA.
The PAs of the difference are found to be $\Delta{\rm PA}_B=-18\fdg5 \pm 2\fdg1$, $\Delta{\rm PA}_V=-27\degr \pm 8\degr$, and $\Delta{\rm PA}_R=-11\fdg6 \pm 1\fdg6$, assuming that the polarization decreased from HIMS to SIMS.
If, instead, the polarization increased, the aforementioned PAs would be shifted by $90\degr$.

Quantifying the change in observed PD and PA acts as an independent method of determining the axis of polarization in the source as this is expected to be only due to the change in intrinsic polarization itself (since IS polarization is taken to be constant). The Hotelling's $T^2$ test performed above would yield the same result as well, since neither statistical nor systematic errors of the IS polarization contribute to the significance of change in polarization. However,
to constrain the intrinsic polarization component more robustly, we next measured the IS polarization.

\begin{table*}
\caption{Average polarization of the field stars.}       
\label{table:avgism}      
\centering          
\begin{tabular}{c c c c c c c}     
\hline\hline       
  &  \multicolumn{2}{c}{$B$} & \multicolumn{2}{c}{$V$} & \multicolumn{2}{c}{$R$} \\ 
\hline
Star  & PD (\%) & PA (deg) & PD (\%) & PA (deg) & PD (\%) & PA (deg) \\
Average 2, 4, 5     & 1.116 $\pm$ 0.001   & 74.19 $\pm$ 0.02 & 1.183 $\pm$ 0.001 & 70.25 $\pm$ 0.02 & 1.197 $\pm$ 0.001 & 70.56 $\pm$ 0.02 \\
Average 2--5    & 1.119 $\pm$ 0.001  & 74.05 $\pm$ 0.02 & 1.231 $\pm$ 0.001 & 68.97 $\pm$ 0.02 & 1.212 $\pm$ 0.001 & 70.10 $\pm$ 0.02 \\
Average 2--9    & 1.122 $\pm$  0.001  & 74.02 $\pm$ 0.02 & 1.226 $\pm$ 0.001 & 69.05 $\pm$ 0.02 & 1.219 $\pm$ 0.001 & 70.23 $\pm$ 0.02 \\
\hline
\end{tabular}
\tablefoot{The full list of field star polarization is given in Table \ref{table:ism}.}
\end{table*}

\begin{figure*}
\centering
\includegraphics[width=0.9\linewidth]{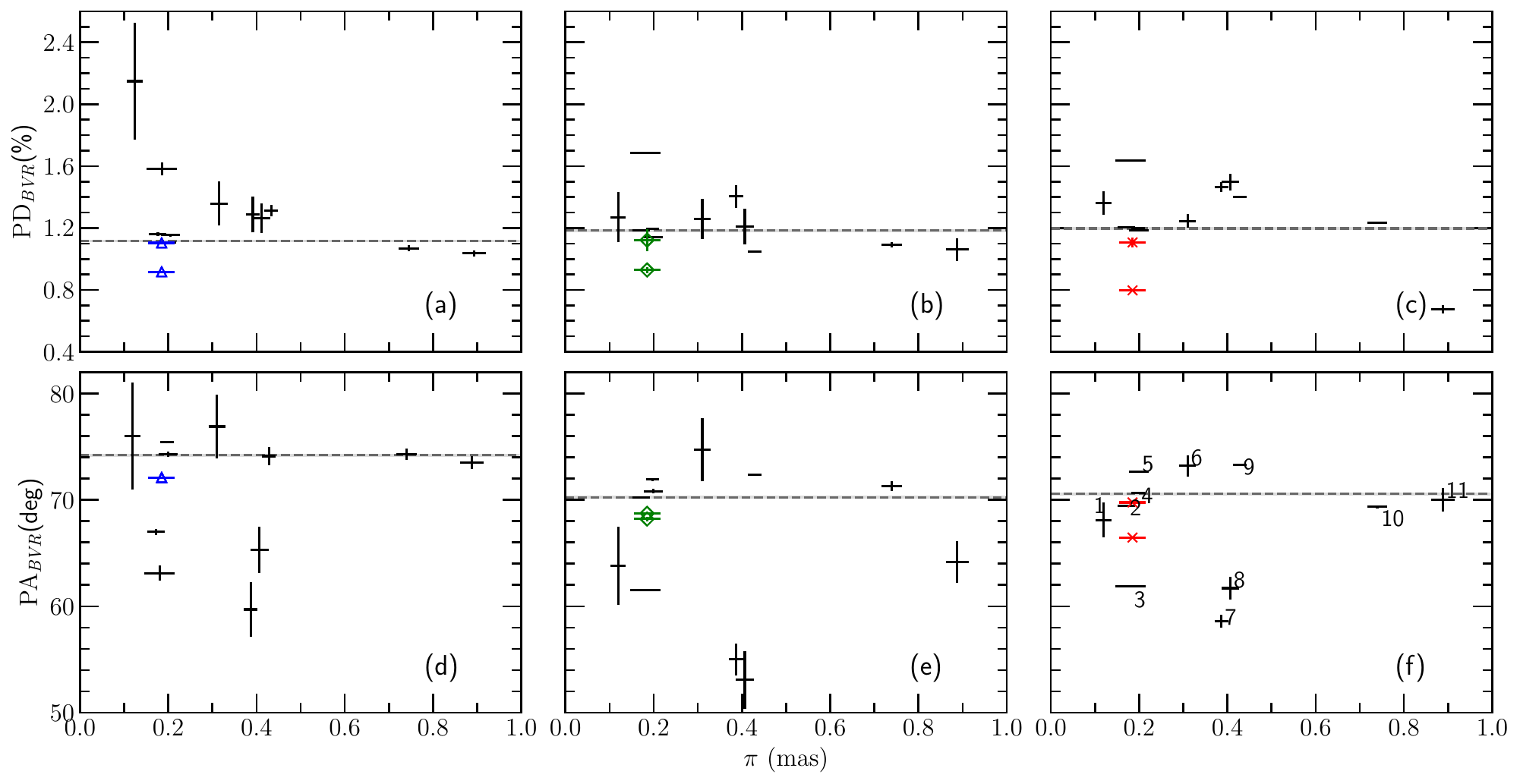}
\caption{Polarization of \source and field stars as a function of {\it Gaia} parallax. PD (top panels) and PA (bottom panels) of the field stars (black crosses, numbers in panel f) and \source (colored symbols) in $B$ (panels a and d), $V$ (panels b and e), and $R$ (panels c and f) bands. The average PD and PA of the source during HIMS and SIMS are shown with colored symbols. The horizontal dashed gray line marks the average polarization of the field stars 2, 4, and 5, with the gray area denoting its $3\sigma$ error.}
\label{fig:PDPAplx}
\end{figure*}

\subsection{Interstellar polarization}

The dust along the line of sight introduces additional IS polarization that must be subtracted from the observed one to recover the intrinsic polarization of the source.
We studied the IS polarization by measuring the polarization of 11 field stars located  within $14\arcmin$ of the source (see Fig.~\ref{fig:sky})  and at a range of distances as determined by \textit{Gaia} parallaxes. 
The list of observed field stars and their polarization properties is given in Table~\ref{table:ism} and shown in Fig.~\ref{fig:PDPAplx}.

The distance  to \source is not well established. \citet{Burridge2025} quote $d=5.5^{+1.4}_{-1.1}$~kpc, while  \citet{MataSanchez2025} give $3.4\pm0.3$~kpc.
These estimates correspond to parallaxes in the range of $\pi\sim$0.15--0.32~mas. 
Thus, we used two options for the field star selection. 
Our primary choice was to select stars at higher distances with $\pi$ around 0.2~mas (i.e., stars 2--5, see Fig.~\ref{fig:sky} and Table~\ref{table:avgism}). 
The secondary option was to use a larger set of stars also covering lower distances with $\pi$ in a wider range (i.e., stars 2--9).

\begin{table*}
\caption{Average intrinsic optical polarization of \source in different spectral states considering two estimates for IS polarization: using field stars 2, 4, and 5, and using field stars 2--9.}            
\label{table:avgintrinsic}      
\centering          
\begin{tabular}{c l  c r c r c r }     
\hline\hline       
 & & \multicolumn{2}{c}{$B$} & \multicolumn{2}{c}{$V$} & \multicolumn{2}{c}{$R$}\\ 
State & MJD & PD (\%) & {PA (deg)} & PD (\%) & {PA (deg)} & PD (\%) & {PA (deg)} \\
\hline
\multicolumn{8}{c}{Average 2, 4, 5} \\
HIMS & 60189--60205 & 0.211 $\pm$ 0.001 & $-$6.6 $\pm$ 0.2 & 0.260 $\pm$ 0.002 &$-14.3\pm 0.2$ & $0.424\pm0.002$ & $-11.7\pm0.1$  \\
SIMS & 60206--60215 & 0.082 $\pm$ 0.002 & 23.9 $\pm$ 0.6 & 0.103 $\pm$ 0.007 & $6.2\pm2.0$ & 0.096 $\pm$ 0.003 & $-$10.0 $\pm$ 0.9 \\
HIMS+SIMS  & 60189--60215 & 0.185 $\pm$ 0.001 & $-4.9 \pm 0.2$ & 0.229 $\pm$ 0.002 & $-12.8\pm0.3$ & $0.345\pm0.002$ & $-11.6\pm0.2$ \\
SS   & 60351--60361 & $<0.33$ & \dots\;\;\; & 0.29 $\pm$ 0.10 & $18\pm10$ & 0.45 $\pm$ 0.05 & 7.8 $\pm$ 3.1  \\
DHS  & 60388--60391 & 0.13 $\pm$ 0.07 & $13 \pm 16$\; & 0.48 $\pm$ 0.09 &  67 $\pm$ 5\,  & 0.09 $\pm$ 0.08 &  $-20 \pm 23$  \\
\hline         
\multicolumn{8}{c}{Average 2--9} \\
HIMS & 60189--60205 & 0.215 $\pm$ 0.001 & $-7.7$ $\pm$ 0.2 & 0.297 $\pm$ 0.002 & $-20.0$ $\pm$ 0.2 & 0.441 $\pm$ 0.002 & $-12.9$ $\pm$ 0.1  \\
SIMS & 60206--60215 & 0.077 $\pm$ 0.002 & 21.4 $\pm$ 0.6 & 0.109 $\pm$ 0.007 & $-12.0$ $\pm$ 1.9 & 0.113 $\pm$ 0.003 & $-15.1$ $\pm$ 0.7  \\
HIMS+SIMS & 60189--60215 & 0.188 $\pm$ 0.001 & $-6.2$ $\pm$ 0.2 &  0.264 $\pm$ 0.002 & $-19.4$ $\pm$ 0.3 & 0.362 $\pm$ 0.002 & $-13.1 $ $\pm$ 0.2  \\
SS & 60351--60361 &  0.12 $\pm$ 0.11  &  $-75 \pm 27$  & 0.26 $\pm$ 0.10 & 12 $\pm$ 11 & 0.45 $\pm$ 0.05 & 6.2 $\pm$ 3.1   \\
DHS & 60388--60391 & 0.13 $\pm$ 0.07 & 12 $\pm$ 16 & 0.44 $\pm$ 0.09 & 70 $\pm$ 6 & 0.12 $\pm$ 0.08 & $-23$ $\pm$ 19   \\
\hline
\end{tabular}
\tablefoot{The PA is given only for measurements where the PD has a significance $>1\sigma$. 
We note that the PA for low-significance detections does not follow a normal distribution \citep{Vinokur1965,Clarke1986,Naghizadeh1993} and the error on the PA is just a formal error from Eq.~\eqref{eq:errPA}. 
The quoted upper limits on the PD are at the $3\sigma$ level. }
\end{table*}

\begin{figure}
\centering
\includegraphics[width=0.9\linewidth]{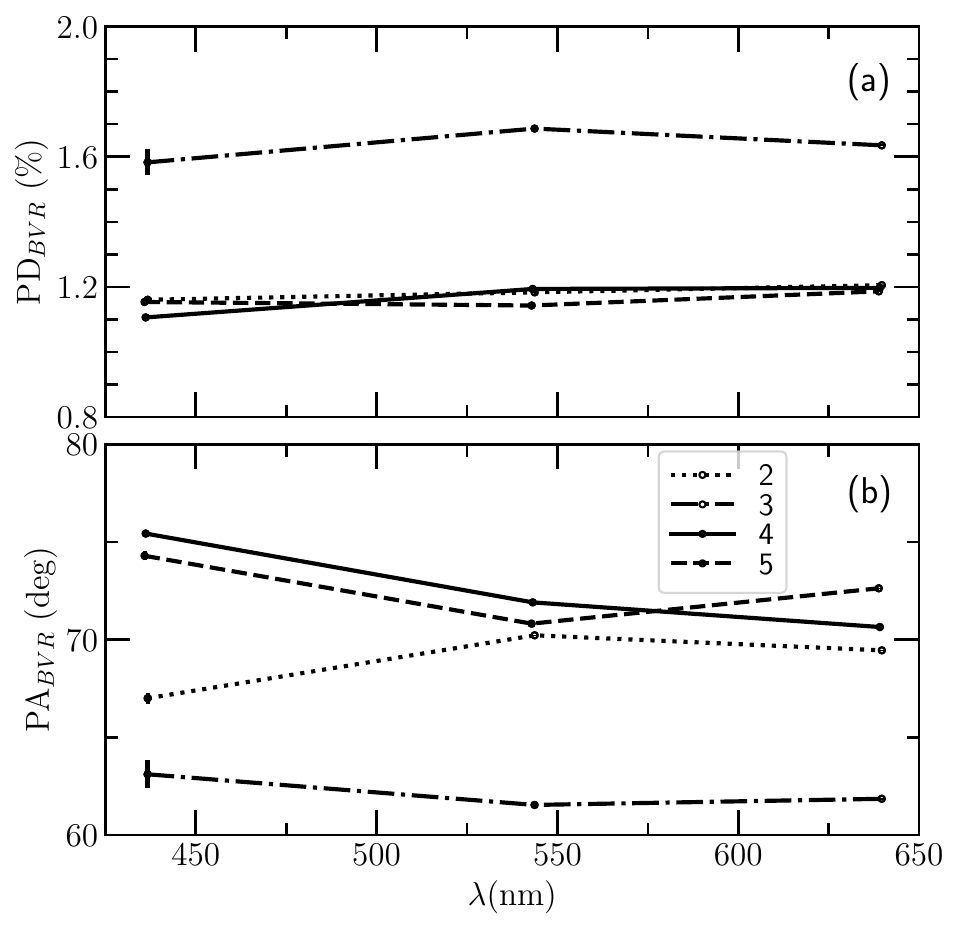}
\caption{Wavelength dependence of PD (panel a) and PA (panel b) of selected field stars.
Star 3 has a noticeably higher PD and different PA, compared to other field stars, which may result from the presence of intrinsic stellar polarization.}
\label{fig:ism_energy}
\end{figure}

We started with the first option. 
To ensure that the selected stars do not exhibit significant intrinsic polarization, we examined the wavelength dependence of their polarization (see Fig.~\ref{fig:ism_energy}).
We found that star \#3 shows PD and PA values that deviate noticeably from those of other stars at comparable distances, suggesting possible intrinsic stellar polarization. 
Therefore, we used the weighted average polarization of stars 2, 4, and~5 (with the weights being inversely proportional to the square of the error on the individual star polarization measurements) to determine the IS component of the optical polarization toward \source.
The PD and PA for that are listed in Table~\ref{table:avgism} and shown as the shaded gray region in Fig.~\ref{fig:PDPAplx}, while the normalized Stokes parameters $q_{\rm is}$ and $u_{\rm is}$ are shown as gray circles in Fig.~\ref{fig:qu_all}.

\begin{figure*}
\centering
\includegraphics[width=0.32\linewidth]{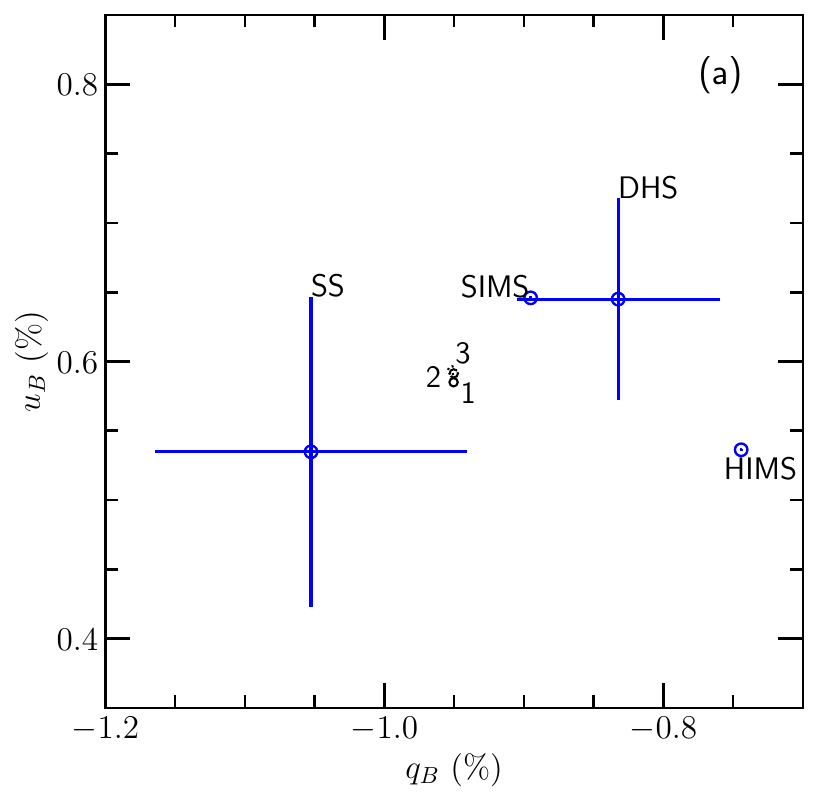}
\includegraphics[width=0.32\linewidth]{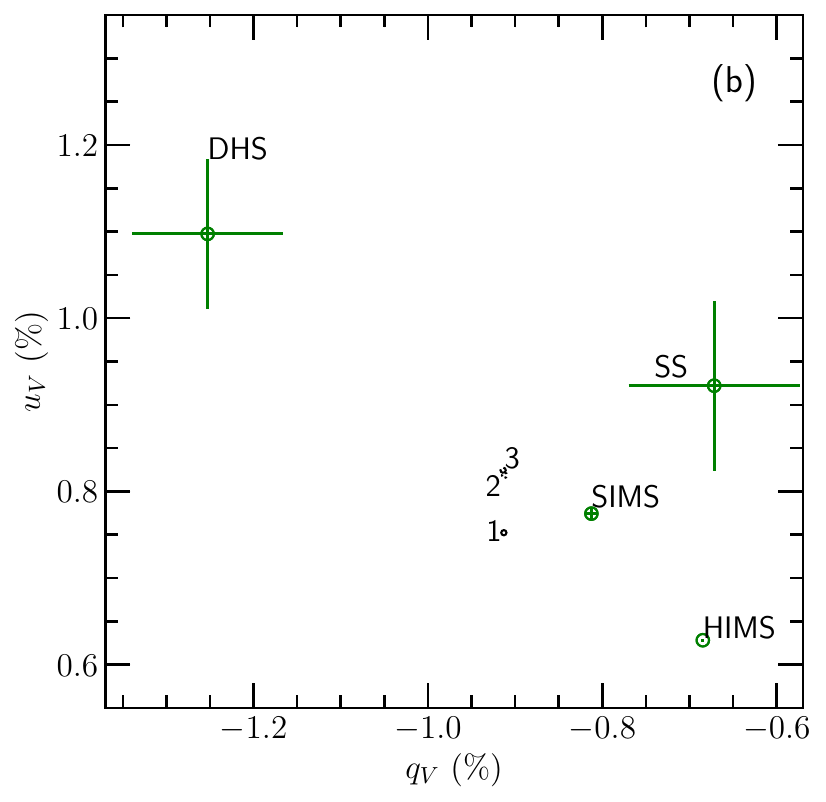}
\includegraphics[width=0.33\linewidth]{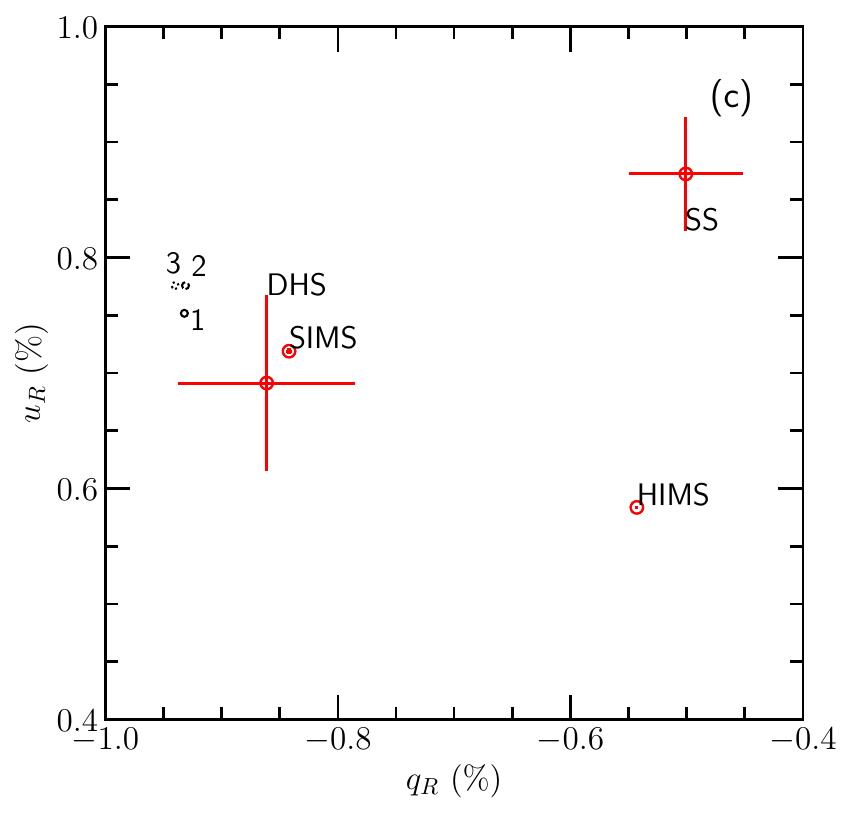}
\caption{Observed Stokes $q$ and $u$ parameters of \source (colored crosses) during HIMS, SIMS, SS, and DHS in (a) $B$, (b) $V$, and (c) $R$ bands. The solid, dashed, and dotted circles (marked as 1, 2, and 3) correspond to the IS polarization estimates using stars 2, 4 and 5, stars 2--5, and stars 2--9, respectively, at a $3\sigma$ error.}
\label{fig:qu_ism_three}
\end{figure*}

We see in Fig.~\ref{fig:sky} that star \#3 is on the other side of the sky compared to stars 2, 4, and 5. 
The difference in its polarization could be due to its position rather than an intrinsic component. 
Thus, we checked the effects of including star \#3 in our IS estimate (Table~\ref{table:avgism}; also shown as dashed circle in Fig.~\ref{fig:qu_ism_three}). 
We see that this increases the IS PD estimate, with the maximum change seen in the $V$ band.

The IS polarization estimate for the second option that includes stars 2--9 is presented in Table~\ref{table:avgism} (also shown as a dotted circle in Fig.~\ref{fig:qu_ism_three}). 
This IS estimate accounts for the variation in the IS polarization from east to west on the sky plane, as well as the dependence of the IS polarization on the distance between the source and the observer. 
This way the estimated IS PD is slightly higher in the $B$ and $R$ bands relative to that for stars 2--5 but lower in the $V$ band. 

Figure~\ref{fig:qu_ism_three} shows on the $q$--$u$ plane the three IS polarization estimates given in Table~\ref{table:avgism} as solid, dashed, and dotted circles of a $3\sigma$ error radius, respectively.  
In all the bands, the IS estimate slightly shifts toward a higher $u$ value for the latter two estimates (largest increase seen in the $V$ band), while the $q$ parameter remains consistent.

\begin{figure*}
\centering
\includegraphics[width=0.9\linewidth]{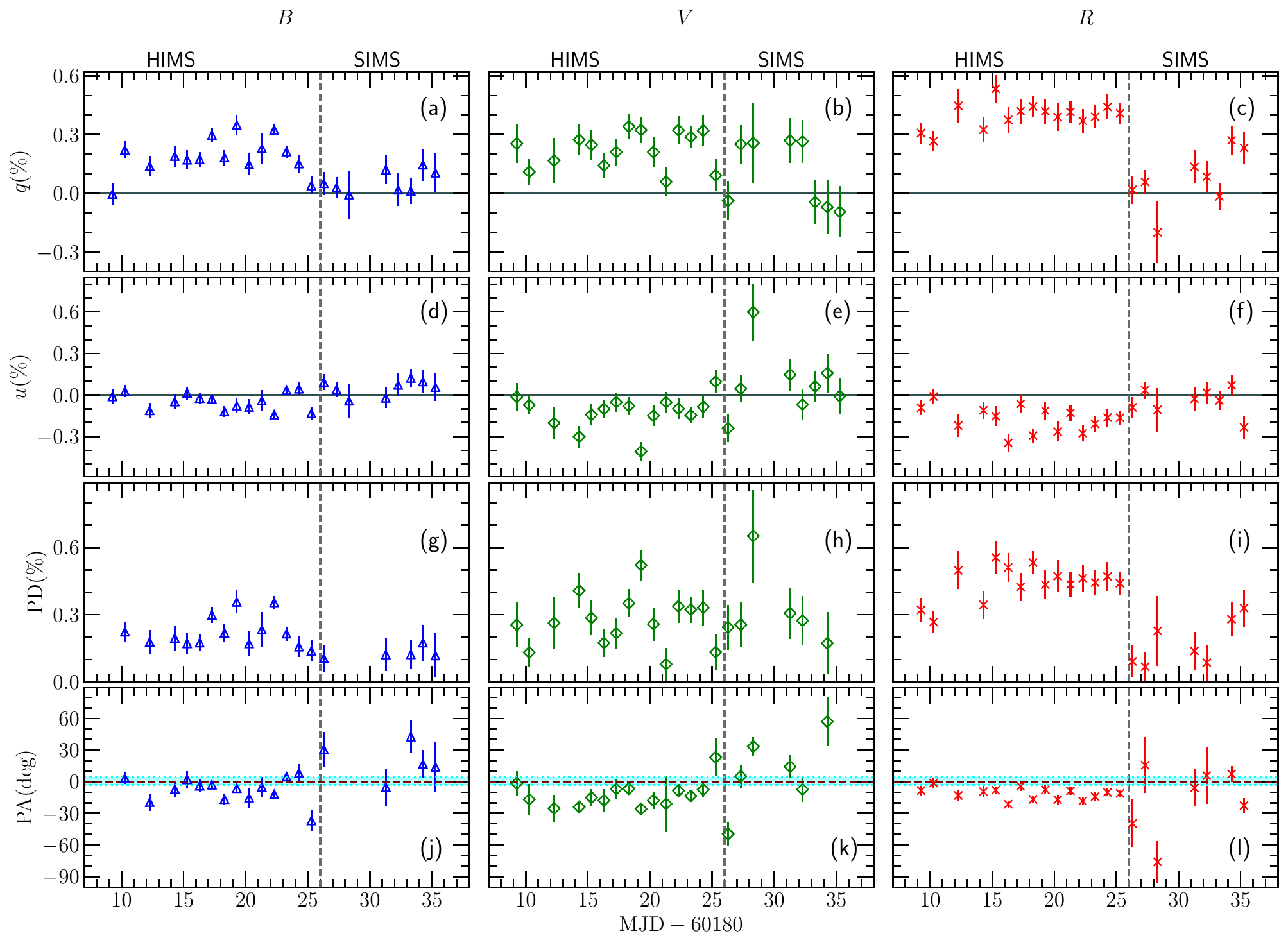}
\caption{Intrinsic (a-c) $q$, (d-f) $u$, (g-i) PD, and (j-l) PA of \source in $B$ (blue triangles; left panels), $V$ (green diamonds; middle panels), and $R$ (red crosses; right panels) bands in the HIMS and SIMS (errors are $1\sigma$) assuming the IS polarization estimate using stars 2, 4, 5. The vertical dashed gray line corresponds to MJD 60206, when the ejections were detected. The gray regions in the top two rows denote the error on IS polarization at the $3\sigma$ level. The horizontal maroon dashed line in the bottom row corresponds to the jet position angle. The shaded cyan area in the bottom row corresponds to the detected X-ray PA.}
\label{fig:PDPA_int}
\end{figure*}

\subsection{Intrinsic polarization}

The intrinsic polarization (subscript ``int'') was calculated by subtracting the averaged Stokes parameters of the IS component from the observed Stokes parameters (subscript ``obs'') of the source:
\begin{equation}
    q_{\rm int} = q_{\rm obs} - q_{\rm is},  \qquad 
    u_{\rm int} = u_{\rm obs} - u_{\rm is} . 
\end{equation} 
We first used the average of stars 2, 4, and 5 as our IS polarization estimate.
The evolution of intrinsic PD and PA with time is shown in Fig.~\ref{fig:PDPA_int} for the whole transition.
The intrinsic polarization values presented here only contain the statistical errors from the observed polarization measurements. To account for the shift in these values due to the error on IS polarization, we show them as gray error bars on the Stokes parameters in Fig.~\ref{fig:PDPA_int} ($q$ and $u$ panels). 
The resulting weighted average values (also accounting for IS polarization estimate errors) are given in Table~\ref{table:avgintrinsic} for HIMS, SIMS, SS, and DHS (values for nightly averages are given in Table~\ref{table:intrinsic}). 
The intrinsic PA are similar to the PA values of the variable component (i.e., $\Delta$PA) estimated in Sect.~\ref{sec:obs_pol} within errors, supporting our choice of the IS polarization.

During HIMS, we find small but statistically significant polarization with PD$_{\rm int}{\approx}0.3\%$.
The PD in the $B$ band is consistently lower than in the other two bands and becomes suppressed during the time associated with the X-ray flare and radio ejections, at the transition to the SIMS.
The change in polarization during the flare is also clearly seen in the $R$ band, where the Stokes $q$ parameter abruptly drops.
In the first week of observations following the flare, the $B$ and $R$-band PD are consistent with zero but begin to recover toward the pre-flare values in the last two observations, leading to the average post-flare PD$_{\rm int}{\approx}0.1\%$.
Interestingly, no apparent change in $V$ band is observed (as is also shown by the Hotelling test in Sect.~\ref{sec:obs_pol}), and the PD$_{\rm int}$ remains at the level ${\sim}0.3\%$ throughout the HIMS.

During HIMS, the PAs in all bands are systematically offset from the X-ray PA=$2\fdg2\pm1\fdg3$ \citep{Veledina2023}, submillimeter PA=$-3\fdg4\pm1\fdg5$ \citep{Vrtilek2023atel}, and the radio PA=$2\degr\pm2\degr$ \citep{Ingram2024}, as well as from the jet position angle of $-0\fdg60\pm0\fdg07$ \citep{Wood2024}, as can be seen in the bottom panels of Fig.~\ref{fig:PDPA_int}. 
The PA of the difference of Stokes parameters, $\Delta{\rm PA}\approx-15\degr$, indicates a misalignment with respect to the X-ray, submillimeter, and radio PAs, as well as from the resolved jet direction.

In the SS, the $V$ and $R$ band PDs are comparable with the HIMS, but the PAs are somewhat different (see Table~\ref{table:avgintrinsic}).
The DHS observations are consistent with the IS values for the obtained error bars, with upper limits on PD ${\lesssim}0.25$\% in the $B$ and $R$ bands. The V-band PD is much higher than it was during the other states, with the PA showing almost a $90\degr$ difference. 

Table~\ref{table:avgintrinsic} also provides the computed intrinsic polarization values for the second option (stars 2--9) of the IS polarization estimate (see also Fig.~\ref{fig:qu_ism_three}).
The polarization values are similar to those obtained with the first IS polarization estimate. 
In all cases, we find a similar effect of depolarization during the SIMS.
Furthermore, we find that in all alternative cases of IS polarization choices, the intrinsic PAs during HIMS differ significantly from the jet direction.
Thus, we conclude that the choice of the IS polarization estimate does not qualitatively affect our conclusions about the average intrinsic polarization and its evolution across the states.

\section{Discussion}
\label{sec:discuss}

\subsection{Origin of optical polarization}

The nature of optical emission during the hard state is being actively investigated using spectral and timing tools \citep{Poutanen2014,UttleyCasella2014}.
Several emission components are considered to be contributing to these wavelengths: thermal emission of the optically thick accretion disk, which can be enhanced due to X-ray irradiation effects, nonthermal synchrotron emission from the accretion flow or from the jet, and the contribution coming from scattering and reprocessing within the slow outflow -- the wind  \citep{Veledina2013,Malzac2014,Munoz-Darias2016}.
The relative role of different components is believed to evolve over the course of the outburst. 
Particularly during transitions to (from) the SS, the role of the nonthermal emission is thought to be decreasing (enhancing), as has been probed by the evolution of OIR colors, as well as the decrease in the flux levels \citep[e.g.,][]{Jain2001,Poutanen2014CMD,Kosenkov2020CMD}.
The aforementioned components are expected to be polarized at different levels \citep[e.g., discussion in][]{Veledina2019}; hence, we expect to see the evolution of the optical PD together with variation in the respective fluxes, most prominent in the red part of the spectrum.

Our observations were performed within the state transition phase, and the optical fluxes followed the expected pattern of a general decrease by ${\sim}0.5$~mag in all bands (see Fig.~\ref{fig:lc_pol}c).
However, the overall behavior of the $BVR$ polarization during HIMS suggests a constant level of PD in all bands, which is most obvious in $R$ (Fig.~\ref{fig:PDPA_int}).
This indicates that either (i) only one component was giving a dominant contribution to $BVR$ fluxes over the transition and preserved its PD despite a changing flux, or (ii)  polarization is produced by scattering of one or several optical components (disk, hot flow, jet) in the wind, whose properties remained unchanged across the HIMS.

The jet contribution to the optical flux and polarization can be estimated using the detection of polarization in the submillimeter band  \citep{Vrtilek2023atel}, with PD=$2.1\pm0.2$\% and $1.9\pm0.2$\% in the direction of the jet (PA=$-3\fdg4\pm1\fdg5$ and $-4\fdg1\pm3\fdg5$), on MJD 60190 and 60191, respectively.
If the jet is the dominant source of optical polarization, then our measurement of PD$_{R}=0.27\pm0.05$\% on MJD 60190.28 (Table~\ref{table:intrinsic}) translates to about 11--16\% contribution of the jet flux in the $R$ band\footnote{Assuming that only the jet contributes to the $R$-band polarization, the $R$-band polarization equals the jet polarization multiplied by the jet’s flux fraction. Using the measured values, the equation becomes $0.27\%$ $=$ jet flux fraction $\times$ $2\%$. Solving this gives a jet-to-total flux ratio of $13.5\%$. Accounting for the $1\sigma$ errors on the measurements, we obtain the range of $11$--$16\%$ flux from the jet in the $R$ band.}.
This fraction is maximal, as it assumes the emission at optical wavelengths is polarized at the same level as in the submillimeter range; the contribution is expected to be smaller if the jet emission at optical wavelengths is polarized at a higher level (as is predicted for the transition from the optically thick to the optically thin synchrotron emission), as compared to the submillimeter.
We find this contribution to be too low to attribute the polarization solely to this component: the disk or hot flow emission, whose fraction is 84--89\%, can produce the observed optical polarization even if they are intrinsically polarized at a sub-percent level ($\lesssim0.3$\%), which can be easily produced.
Hence, in order not to overproduce the intrinsic optical polarization in HIMS, the other components should have polarization $\ll0.1$\%, which may be difficult to organize.
Thus, we consider that the scenario with dominant jet contribution to the observed optical polarization is not a viable option.

Another possibility is the presence of two (or more) polarized components whose polarizations partially cancel each other out. 
This case may arise, for example, from the jet synchrotron emission produced in the predominantly toroidal field (i.e., with the PA along the jet axis, as is observed) and the disk emission polarized according to the pure electron-scattering atmosphere law \citep{Cha60,Sob63} with the PA orthogonal to the jet axis.
The joint contribution of the two components then results in partial depolarization.
The resulting PD is expected to be sensitive to the relative fraction of these components in the total spectrum and to vary with time if one of the components changes flux.
However, the observed drop in optical flux by ${\sim}0.5$~mag was not accompanied by any detectable changes of intrinsic polarization in the HIMS.
Furthermore, the abrupt drop of PD at discrete ejections down to zero requires fine-tuning of the relative contribution of components, which is difficult to organize, especially on a timescale of several days.
We conclude that the option of two orthogonally polarized components is also not viable.

Next, we consider the optically thick disk as the dominant source of optical polarization, as well as optical emission (${\gtrsim}85$\%, according to the estimate above).
For the pure electron scattering atmosphere model \citep{Cha60}, the observed polarization translates to the disk inclination of ${\approx}25\degr$, in line with the low-to-moderate inclination inferred from other methods \citep{MataSanchez2025,Wood2024}.
However, the requirement of the dominance of this component throughout the HIMS faces problems with the observed evolution of the spectral energy distribution and the detection of sub-second quasi-periodic oscillations in OIR light \citep{Vincentelli2025}, which cannot be produced at the distances relevant to the scales of OIR emission in optically thick accretion disks.
This scenario also faces problems with the drop of intrinsic polarization at the time of ejections, as it is not clear what could cause the substantial depolarization.

The synchrotron emission of the hot flow can dominate the optical emission and also produce quasi-periodic oscillations \citep{Veledina2013QPO}.
The decline of the OIR emission during the hard to soft state transition was previously linked to the contraction of the hot flow and its replacement with the optically thick disk \citep{Poutanen2014CMD,Kosenkov2020CMD}; hence, the constant PD level during HIMS could be associated with the persistent dominance of the hot flow component in the optical spectra.
The abrupt drop of polarization at the transition to SIMS may then be associated with the collapse of the flow, an emergence of another unpolarized component, or sudden disordering of the magnetic field within the flow.
The prevailing direction and level of ordering of the magnetic field within the flow are uncertain; in order to produce the PA aligned with the jet axis during HIMS, the toroidal field should dominate.
However, the systematic difference between the intrinsic PA in the HIMS from the X-ray and submillimeter PA poses a severe problem for this scenario.

Finally, we considered the possibility that the optical polarization is produced by scattering of a fraction of emission coming from the inner regions (hot flow or jet) in the optically thin outflow -- the wind. 
Scattering in a wind was recently proposed to explain the high levels of X-ray polarization observed in X-ray binaries \citep{Nitindala2025}.
However, it can also be the source of polarization in the optical bands, where light from the hot flow or the base of the jet is scattered.
The PD of the radiation scattered in the wind  generally depends on the inclination of the source, but also on the opacity of the scattering material, the characteristic wind opening angle, and the angular distribution of the incident emission, although the dependence on wind parameters is rather weak for low inclinations.
Figure~\ref{fig:lc} shows that the spectroscopic detections of the disk winds were simultaneous with our measurements of optical polarization, indicating that the winds were indeed present during the HIMS.
Although no spectroscopic observations were obtained near the time of the radio ejections, winds were detected both several days afterward and during the subsequent transition into the SIMS.
Whether the wind properties changed around the time of the radio ejections, potentially contributing to the observed reduction in polarization, remains to be determined.

The wind scattering scenario may offer a viable explanation for the mismatch of the optical PA from the X-ray PA and the jet axis.
The winds are launched at high distances from the central source, and hence may inherit the axis of symmetry of the outer parts of the disk, which may align with the orbital axis.
The X-ray and submillimeter polarization were instead found to be aligned with the jet axis, which is also thought to coincide with the axis of the BH spin.
The mismatch between the optical and X-ray PAs can then be interpreted as the signature of misalignment between the BH and orbital axes in \source, similar to MAXI J1820+070 \citep{Poutanen2022}.

\subsection{Comparison to other BH X-ray binaries}

The evolution of the optical polarimetric signatures of \source is in many ways similar to those detected in the outburst of MAXI J1820+070 \citep{Veledina2019,Kosenkov2020,Poutanen2022}.
Optical polarization of \source during HIMS (PD${\sim}0.3$\%) is at the same sub-percent level as that detected in the rising phase of MAXI J1820+070 (PD${\sim}0.8$\%).
The PD was found to increase with the progression of the outburst in MAXI J1820+070; however, we did not see any significant PD changes in \source during HIMS, and instead detected a drop in polarization after the ejections. This might be related to the difference of inclinations ($\sim$75\degr\ for MAXI J1820+070), or be partially caused by the somewhat different coverage of the hardness-intensity diagram.

The optical PA is well aligned with the jet direction in MAXI J1820+070, and can be related to the dominant role of synchrotron emission and polarization either from the hot flow or from the jet, as the average PA of both components is expected to be aligned with the jet axis.
For \source, the systematic difference with respect to the X-ray PA might be caused by the different dominant optical polarization mechanism; namely, by the scattering in the wind, as was discussed above, which traces the (misaligned) orbital axis.
The presence of a misalignment is also expected owing to the detection of quasi-periodic oscillations in the system \citep{Vincentelli2025}, which are thought to be related to the Lense-Thirring precession of the hot flow \citep{Fragile2007,Ingram2009}.
Misalignment can be expected for systems where the newly born BH was produced in an asymmetric supernova explosion, which is also thought to lead to a substantial natal kick.
The natal kick velocity in \source was found to be  $220^{+30}_{-40}$~km~s$^{-1}$ \citep{MataSanchez2025}, which is higher than the one inferred for MAXI~J1820+070 and among the highest in all BH X-ray binaries \citep{Atri2019}.
Dedicated polarimetric observations during (near-)quiescence are needed in order to verify the misalignment in the system and to have a more precise measurement of the difference in position angles of the orbital and jet axes.

\section{Summary}
\label{sec:summary}

We present the first optical polarization measurements of \source throughout its 2023-2024 outburst. 
The observed polarization of ${\sim}1$\% is detected in the $BVR$ bands for all nights during HIMS and SIMS, with confidence levels exceeding $7\sigma$.
Lower fluxes during SS and DHS cause a reduction in the signal-to-noise level, leading to a decrease in significance.
For the first time in a BH X-ray binary, we detect a significant change in optical polarization coincident with the discrete radio ejections, which are associated with the HIMS-SIMS transition.

To accurately account for the IS polarization, we analyzed nearby field stars. 
We identified stars at similar distances to the source at a small angular separation and subtracted their average normalized Stokes parameters from the observed values.
After this correction, we find that the intrinsic PD remained stable at PD$\approx$0.3\% throughout the HIMS, but dropped abruptly at the transition to SIMS, becoming consistent with zero within the uncertainties for about a week after the transition.

During the HIMS, the source flux has declined by approximately 0.5~mag; hence, the observed stability of PD in this state imposes strong constraints on the plausible polarization mechanisms.
We considered polarized emission from the jet, hot accretion flow, and optically thick disk, but found that none of these can fully explain the observed behavior.
Our preferred interpretation is that the optical polarization during the HIMS arises from scattering in an optically thin disk wind.

The intrinsic PA is systematically offset from the X-ray, submillimeter, and radio PAs by about $-15\degr$, matching the PA of the variable component, $\Delta{\rm PA}$, observed across the HIMS--SIMS transition.
We interpret this offset as evidence of a misalignment between the BH spin and orbital axis, which could be tested by targeted polarimetric observations of \source near quiescence.

\begin{acknowledgements}
This research has been supported by the Magnus Ehrnrooth Foundation (APN) and the Academy of Finland grant 355672 (AV).
Nordita is supported in part by NordForsk. The DIPOL-2 was built in cooperation by the University of Turku,
Finland, and the Kiepenheuer Institut fuer Sonnenphysik, Germany, with support from the Leibniz Association grant SAW-2011-KIS-7. We are grateful to
the Institute for Astronomy, University of Hawaii for the observing time allocated for us on the T60 telescope. 
The data collected using DIPol-UF were based on observations made with the Nordic Optical Telescope, owned in collaboration by the University of Turku and Aarhus University, and operated jointly by Aarhus University, the University of Turku and the University of Oslo, representing Denmark, Finland and Norway, the University of Iceland and Stockholm University at the Observatorio del Roque de los Muchachos, La Palma, Spain, of the Instituto de Astrofisica de Canarias. The NOT data were obtained under program ID P71-015.
We acknowledge with thanks the variable star observations from the AAVSO International Database contributed by observers worldwide and used in this research. 
This work has made use of data from the European Space Agency (ESA) mission
{\it Gaia} (\url{https://www.cosmos.esa.int/gaia}), processed by the {\it Gaia}
Data Processing and Analysis Consortium (DPAC,
\url{https://www.cosmos.esa.int/web/gaia/dpac/consortium}). 
Funding for the DPAC has been provided by national institutions, in particular the institutions participating in the {\it Gaia} Multilateral Agreement. 
This research has made use of \textit{"Aladin sky atlas"} developed at CDS, Strasbourg Observatory, France  \citep{Aladinv3}.
\end{acknowledgements} 

%
%
\bibliographystyle{aa} 
\bibliography{references.bib} 

\begin{appendix}
\onecolumn
\begin{table*}
\section{Optical polarization data} 
\caption{Observed optical polarization of \source in the different spectral states.}             
\label{table:obs}      
\centering          
\begin{tabular}{ccccccc}       
\hline\hline       
  & \multicolumn{2}{c}{$B$} & \multicolumn{2}{c}{$V$} & \multicolumn{2}{c}{$R$}\\ 
MJD & PD (\%) & PA (deg) & PD (\%) & PA (deg) & PD (\%) & PA (deg) \\
\hline
\multicolumn{7}{c}{HIMS} \\
60189.27609 & 1.11 $\pm$ 0.05 & 74.5 $\pm$ 1.4 & 0.99 $\pm$ 0.10 & 65.9 $\pm$ 2.9 & 0.91 $\pm$ 0.05 & 66.7 $\pm$ 1.7 \\ 
    60190.28030 & 0.95 $\pm$ 0.04 & 70.0 $\pm$ 1.3 & 1.05 $\pm$ 0.07 & 69.9 $\pm$ 1.8 & 1.00 $\pm$ 0.05 & 66.0 $\pm$ 1.5 \\ 
60192.27699 & 0.94 $\pm$ 0.05 & 74.9 $\pm$ 1.6 & 0.93 $\pm$ 0.12 & 71.8 $\pm$ 3.6 & 0.72 $\pm$ 0.08 & 66.2 $\pm$ 3.4  \\
    60194.29840 & 0.93 $\pm$ 0.05 & 72.4 $\pm$ 1.7 & 0.78 $\pm$ 0.08 & 72.4 $\pm$ 2.9 & 0.88 $\pm$ 0.06 & 66.7 $\pm$ 2.0  \\
60195.28800  & 0.98 $\pm$ 0.05 & 71.3 $\pm$ 1.4 & 0.90 $\pm$ 0.08 & 68.8 $\pm$ 2.5 & 0.72 $\pm$ 0.07 & 61.8 $\pm$ 2.8 \\
    60196.29440  & 0.96 $\pm$ 0.04 & 72.1 $\pm$ 1.2 & 1.01 $\pm$ 0.06 & 69.9 $\pm$ 1.8 & 0.69 $\pm$ 0.06 & 72.0 $\pm$ 2.7 \\
60197.28450  & 0.86 $\pm$ 0.04 & 69.8 $\pm$ 1.2 & 0.99 $\pm$ 0.07 & 67.5 $\pm$ 2.0 & 0.86 $\pm$ 0.06 & 63.4 $\pm$ 2.1 \\
    60198.29669  & 0.90 $\pm$ 0.04 & 74.4 $\pm$ 1.3 & 0.88 $\pm$ 0.06 & 65.2 $\pm$ 2.1 & 0.67 $\pm$ 0.05 & 68.4 $\pm$ 2.2 \\
60199.28100  & 0.79 $\pm$ 0.05 & 69.9 $\pm$ 1.9 & 0.68 $\pm$ 0.07 & 74.9 $\pm$ 2.8 & 0.82 $\pm$ 0.06 & 64.4 $\pm$ 2.2 \\
    60200.27389  & 0.95 $\pm$ 0.06 & 74.1 $\pm$ 1.7 & 0.93 $\pm$ 0.08 & 69.7 $\pm$ 2.3 & 0.73 $\pm$ 0.07 & 69.0 $\pm$ 2.8 \\
60201.29749  & 0.90 $\pm$ 0.08 & 71.5 $\pm$ 2.4 & 1.10 $\pm$ 0.07 & 70.3 $\pm$ 1.9 & 0.81 $\pm$ 0.06 & 64.8 $\pm$ 2.0 \\
    60202.29339  & 0.77 $\pm$ 0.03 & 72.4 $\pm$ 1.1 & 0.88 $\pm$ 0.07 & 66.1 $\pm$ 2.4 & 0.74 $\pm$ 0.06 & 69.9 $\pm$ 2.3 \\
60203.28620  & 0.96 $\pm$ 0.03 & 70.0 $\pm$ 1.0 & 0.87 $\pm$ 0.06 & 68.0 $\pm$ 1.9 & 0.77 $\pm$ 0.06 & 67.5 $\pm$ 2.2 \\
    60204.28629  & 1.02 $\pm$ 0.05 & 70.9 $\pm$ 1.3 & 0.89 $\pm$ 0.08 & 65.8 $\pm$ 2.6 & 0.77 $\pm$ 0.06 & 64.9 $\pm$ 2.4 \\
60205.28880 & 1.02 $\pm$ 0.05 & 76.8 $\pm$ 1.3 & 1.18 $\pm$ 0.08 & 67.0 $\pm$ 2.0 & 0.79 $\pm$ 0.05 & 65.9 $\pm$ 1.9 \\
\multicolumn{7}{c}{SIMS} \\
    60206.29199 & 1.13 $\pm$ 0.06 & 71.5 $\pm$ 1.5 & 1.08 $\pm$ 0.10 & 75.9 $\pm$ 2.6 & 1.13 $\pm$ 0.07 & 72.1 $\pm$ 1.8 \\
60207.31030  & 1.11 $\pm$ 0.06 & 73.0 $\pm$ 1.4 & 1.04 $\pm$ 0.10 & 64.8 $\pm$ 2.7 & 1.18 $\pm$ 0.06 & 69.0 $\pm$ 1.5 \\
    60208.31730  & 1.10 $\pm$ 0.12 & 75.3 $\pm$ 3.2 & 1.50 $\pm$ 0.21 & 57.9 $\pm$ 3.9 & 1.30 $\pm$ 0.16 & 75.2 $\pm$ 3.4 \\
60211.28109  & 1.00 $\pm$ 0.07 & 72.9 $\pm$ 2.1 & 1.11 $\pm$ 0.11 & 62.8 $\pm$ 2.9 & 1.08 $\pm$ 0.08 & 68.9 $\pm$ 2.2 \\
    60212.27909 & 1.14 $\pm$ 0.08 & 72.4 $\pm$ 2.1 & 0.94 $\pm$ 0.11  & 66.8 $\pm$ 3.4 & 1.14 $\pm$ 0.08 & 68.9 $\pm$ 2.0 \\
60213.29499  & 1.18 $\pm$ 0.07 & 71.5 $\pm$ 1.6 & 1.26 $\pm$ 0.11 & 69.8 $\pm$ 2.6 & 1.19 $\pm$ 0.07 & 71.6 $\pm$ 1.6 \\
    60214.27279  & 1.05 $\pm$ 0.08 & 69.9 $\pm$ 2.2 & 1.34 $\pm$ 0.14 & 68.6 $\pm$ 3.0 & 1.06 $\pm$ 0.08 & 64.4 $\pm$ 2.0 \\
60215.27840  & 1.06 $\pm$ 0.10 & 71.4 $\pm$ 2.7 & 1.25 $\pm$ 0.13 & 71.8 $\pm$ 3.0 & 0.87 $\pm$ 0.08 & 71.8 $\pm$ 2.8 \\
\multicolumn{7}{c}{SS} \\
60351.65640 & 2.65 $\pm$ 0.57 & 71 $\pm$ 6 & 1.79 $\pm$ 0.89 & 38 $\pm$ 13 & 1.40 $\pm$ 0.52 & 92 $\pm$ 10 \\
60352.63549  & 1.00 $\pm$ 0.41 & 74 $\pm$ 11 & 1.05 $\pm$ 0.56 & 11 $\pm$ 70 & 0.95 $\pm$ 0.36 & 65 $\pm$ 10 \\
60355.63249  & 1.09 $\pm$ 0.35 & 96 $\pm$ 9 & 1.14 $\pm$ 0.48 & 39 $\pm$ 12 & 1.03 $\pm$ 0.21 & 48$\pm$ 6 \\
60359.62550  & 1.31 $\pm$ 0.51 & 66 $\pm$ 11 & 1.59 $\pm$ 0.56 & 84 $\pm$ 10 & 0.48 $\pm$ 0.37 & 75 $\pm$ 19  \\
60360.62740  & 1.26 $\pm$ 0.59 & 62 $\pm$ 13 & 1.21 $\pm$ 0.28 & 66 $\pm$ 7 & 1.21 $\pm$ 0.28 & 66 $\pm$ 7 \\
60361.60220  & $< 4.0$ & \dots  & 2.51 $\pm$ 1.13 & 53 $\pm$ 12 & 1.69 $\pm$ 0.35 & 57 $\pm$ 6 \\
\multicolumn{7}{c}{DHS} \\
60388.59800  & 1.48 $\pm$ 0.32 & 67 $\pm$ 6 & 1.91 $\pm$ 0.32 & 74 $\pm$ 5 & 1.59 $\pm$ 0.36 & 70 $\pm$ 6 \\
60390.55810  & 0.88 $\pm$ 0.36 & 78 $\pm$ 11 & 1.27 $\pm$ 0.43 & 71 $\pm$ 9 & 1.34 $\pm$ 0.33 & 85 $\pm$ 7 \\
60391.55979  & $< 1.6$ & \dots & 1.87 $\pm$ 0.52 & 56 $\pm$ 8 & 0.87 $\pm$ 0.28 & 55 $\pm$ 9 \\
\hline    
\end{tabular} 
\tablefoot{Upper limits on $3\sigma$ level are provided for detections with a significance below $1 \sigma$.}
\end{table*}

\begin{table*}
\caption{Polarimetric data for the field stars.}        
\label{table:ism}      
\centering          
\begin{tabular}{c c c c c c c c}     
\hline\hline       
  & & \multicolumn{2}{c}{$B$} & \multicolumn{2}{c}{$V$} & \multicolumn{2}{c}{$R$} \\ 
Star & $\pi$ (mas) & PD (\%) & PA (deg) & PD (\%) & PA (deg) & PD (\%) & PA (deg) \\
\hline
1 & 0.12 $\pm$ 0.02 & 2.15 $\pm$ 0.38 & 76.0 $\pm$ 5.0 & 1.27 $\pm$ 0.16 & 63.8 $\pm$ 3.6 & 1.36 $\pm$ 0.08 & 68.1 $\pm$ 1.6  \\
    2 & 0.17 $\pm$ 0.02 & 1.16 $\pm$ 0.01 & 67.0 $\pm$ 0.3 & 1.183 $\pm$ 0.001 & 70.23 $\pm$ 0.04 & 1.205 $\pm$ 0.001 & 69.46 $\pm$ 0.03  \\
3 & 0.18 $\pm$ 0.03 & 1.58 $\pm$ 0.04 & 63.1 $\pm$ 0.7 & 1.686 $\pm$ 0.002 & 61.54 $\pm$ 0.03 & 1.635 $\pm$ 0.003 & 61.85 $\pm$ 0.06    \\
   4 & 0.20 $\pm$ 0.02 & 1.106 $\pm$ 0.005 & 75.4 $\pm$ 0.1 & 1.194 $\pm$ 0.006 & 71.9 $\pm$ 0.2 & 1.197 $\pm$ 0.001 & 70.65 $\pm$ 0.03  \\
5 & 0.20 $\pm$ 0.02 & 1.154 $\pm$ 0.009 & 74.3 $\pm$ 0.2 & 1.143 $\pm$ 0.008 & 70.8 $\pm$ 0.2 & 1.186 $\pm$ 0.002 & 72.63 $\pm$ 0.04  \\
    6 & 0.31 $\pm$ 0.02 & 1.36 $\pm$ 0.14 & 76.9 $\pm$ 3.0 & 1.26 $\pm$ 0.13  & 74.7 $\pm$ 2.9 & 1.25 $\pm$ 0.04 & 73.2 $\pm$ 1.0  \\
7 & 0.39 $\pm$ 0.02 & 1.29 $\pm$ 0.12 & 59.7 $\pm$ 2.6 & 1.40 $\pm$ 0.07  & 55.0 $\pm$ 1.5 & 1.46 $\pm$ 0.03 & 58.6 $\pm$ 0.6  \\
    8 & 0.41 $\pm$ 0.02 & 1.26 $\pm$ 0.09 & 65.3 $\pm$ 2.1 & 1.21 $\pm$ 0.11  & 53.1 $\pm$ 2.7 & 1.50 $\pm$ 0.06 & 61.7 $\pm$ 1.1  \\
9 & 0.43 $\pm$ 0.02 & 1.31 $\pm$ 0.04 & 74.1 $\pm$ 0.8 & 1.046 $\pm$ 0.004 & 72.4 $\pm$ 0.1 & 1.400 $\pm$ 0.003 & 73.31 $\pm$ 0.07  \\
    10 & 0.74 $\pm$ 0.02 & 1.07 $\pm$ 0.02 & 74.3 $\pm$ 0.5 & 1.09 $\pm$ 0.02 & 71.3 $\pm$ 0.4 & 1.234 $\pm$ 0.006 & 69.4 $\pm$ 0.1   \\
11 & 0.89 $\pm$ 0.03 & 1.04 $\pm$ 0.02 & 73.5 $\pm$ 0.6 & 1.06 $\pm$ 0.07 & 64.2 $\pm$ 2.0 & 0.67 $\pm$ 0.03 & 70.0 $\pm$ 1.1  \\ 
\hline

\end{tabular}
\tablefoot{$\pi$ is the parallax to the stars in milliarcseconds. 
In this work, we considered two different distance estimates to \source of $5.5^{+1.4}_{-1.1}$~kpc from \citet{Burridge2025} and $3.4\pm0.3$~kpc from \citet{MataSanchez2025}, which correspond to $\pi=0.18\pm0.04$ and $0.30\pm0.03$~mas, respectively.}
\end{table*}

\begin{table*}
\caption{Intrinsic polarization of \source.}            
\label{table:intrinsic}      
\centering          
\begin{tabular}{l  c r c r c r } 
\hline\hline       
  & \multicolumn{2}{c}{$B$} & \multicolumn{2}{c}{$V$} & \multicolumn{2}{c}{$R$}\\ 
MJD & PD (\%) & PA (deg) & PD (\%) & PA (deg) & PD (\%) & PA (deg) \\
\hline
\multicolumn{7}{c}{HIMS} \\
60189.27609 & $< 0.15$ & \dots & 0.25 $\pm$ 0.10 & $-2$ $\pm$ 11 & 0.32 $\pm$ 0.05 & $-8.4$ $\pm$ 4.8 \\
60190.28030 & 0.22 $\pm$ 0.04 & 3.4 $\pm$ 5.7 & 0.13 $\pm$ 0.07 & $-17$ $\pm$ $15$ & 0.27 $\pm$ 0.05 & $-1.2$ $\pm$ 5.4 \\
60192.27699 & 0.18 $\pm$ 0.05 & $-19.5$ $\pm$ 8.4 & 0.26 $\pm$ 0.12 & $-25$ $\pm$ 13 & 0.50 $\pm$ 0.08 & $-13.1$ $\pm$ 4.9 \\
60194.29840 & 0.20 $\pm$ 0.05 & $-7.2$ $\pm$ 8.0 & 0.41 $\pm$ 0.08 & $-23.9$ $\pm$ 5.5 & 0.34 $\pm$ 0.06 & $-9.5$ $\pm$ 5.2 \\
60195.28800 & 0.17 $\pm$ 0.05 & 2.1 $\pm$ 8.0 & 0.29 $\pm$ 0.08 & $-15.0$ $\pm$ 7.8 & 0.56 $\pm$ 0.07 & $-8.0$ $\pm$ 3.7 \\
60196.29440 & 0.17 $\pm$ 0.04 & $-4.0$ $\pm$ 6.3 & 0.17 $\pm$ 0.06 & $-18$ $\pm$ 10 & 0.51 $\pm$ 0.06 & $-21.4$ $\pm$ 3.6 \\
60197.28450 & 0.30 $\pm$ 0.04 & $-2.9$ $\pm$ 3.4 & 0.22 $\pm$ 0.07 & $-6.9$ $\pm$ 9.0 & 0.42 $\pm$ 0.06 & $-4.4$ $\pm$ 4.3 \\
60198.29669 & 0.22 $\pm$ 0.04 & $-16.7$ $\pm$ 5.2 & 0.35 $\pm$ 0.06 & $-6.6$ $\pm$ 5.2 & 0.53 $\pm$ 0.05 & $-16.8$ $\pm$ 2.8 \\
60199.28100 & 0.36 $\pm$ 0.05 & $-6.3$ $\pm$ 4.1 & 0.52 $\pm$ 0.07 & $-25.8$ $\pm$ 3.7 & 0.43 $\pm$ 0.06 & $-7.5$ $\pm$ 4.2 \\
60200.27389 & 0.17 $\pm$ 0.06 & $-15$ $\pm$ 9 & 0.26 $\pm$ 0.08 & $-17.7$ $\pm$ 8.3 & 0.47 $\pm$ 0.07 & $-17.1$ $\pm$ 4.4 \\
60201.29749 & 0.23 $\pm$ 0.08 & $-5$ $\pm$ 10 & 0.08 $\pm$ 0.07 & $-21$ $\pm$ 27 & 0.44 $\pm$ 0.06 & $-8.6$ $\pm$ 3.7 \\
60202.29339 & 0.35 $\pm$ 0.03 & $-11.9$ $\pm$ 2.4 & 0.34 $\pm$ 0.07 & $-8.6$ $\pm$ 6.3 & 0.46 $\pm$ 0.06 & $-18.4$ $\pm$ 3.7 \\
60203.28620 & 0.22 $\pm$ 0.03 & 4.7 $\pm$ 4.3 & 0.32 $\pm$ 0.06 & $-13.5$ $\pm$ 5.2 & 0.44 $\pm$ 0.06 & $-14.1$ $\pm$ 3.7 \\
60204.28629 & 0.16 $\pm$ 0.05 & 8.2 $\pm$ 8.3 & 0.33 $\pm$ 0.08 & $-7.5$ $\pm$ 6.9 & 0.47 $\pm$ 0.06 & $-10.2$ $\pm$ 3.8 \\
60205.28880 & 0.14 $\pm$ 0.05 & $-37$ $\pm$ 10 & 0.13 $\pm$ 0.08 & 23 $\pm$ 18 & 0.44 $\pm$ 0.05 & $-11.0$ $\pm$ 3.4 \\
\multicolumn{7}{c}{SIMS} \\
60206.29199 & 0.11 $\pm$ 0.06 & 31 $\pm$ 16 & 0.24 $\pm$ 0.10 & $-50$ $\pm$ 12 & 0.09 $\pm$ 0.07 & $-40$ $\pm$ 23 \\
60207.31030 & $< 0.18$ & \dots & 0.26 $\pm$ 0.10 & 5 $\pm$ 11 & 0.07 $\pm$ 0.06 & 16 $\pm$ 27 \\
60208.31730 & $< 0.36$ & \dots & 0.65 $\pm$ 0.21 & 33.4 $\pm$ 9.1 & 0.23 $\pm$ 0.16 & $-76$ $\pm$ 20 \\
60211.28109 & 0.12 $\pm$ 0.07 & $-5$ $\pm$ 18 & 0.31 $\pm$ 0.11 & 14 $\pm$ 11 & 0.14 $\pm$ 0.08 & $-6$ $\pm$ 18 \\
60212.27909 & $< 0.24$ & \dots & 0.27 $\pm$ 0.11 & $-7$ $\pm$ 12 & 0.09 $\pm$ 0.08 & 6 $\pm$ 27 \\
60213.29499 & 0.12 $\pm$ 0.07 & 43 $\pm$ 15 & $< 0.33$ & \dots & $< 0.21$ & \dots \\
60214.27279 & 0.17 $\pm$ 0.08 & 17 $\pm$ 13 & 0.17 $\pm$ 0.14 & 57 $\pm$ 23 & 0.28 $\pm$ 0.08 & 7.3 $\pm$ 7.7 \\
60215.27840 & 0.12 $\pm$ 0.10 & 14 $\pm$ 24 & $< 0.39$ & \dots & 0.33 $\pm$ 0.08 & $-22.6$ $\pm$ 7.3 \\
\multicolumn{7}{c}{SS} \\
60352.14600 & 0.45 $\pm$ 0.20 & 67 $\pm$ 13 & 0.52 $\pm$ 0.28 & 13 $\pm$ 15 & 0.29 $\pm$ 0.12 & $-38$ $\pm$ 12 \\
60357.62900 & 0.43 $\pm$ 0.12 & $-48.8$ $\pm$ 8.2 & 0.37 $\pm$ 0.27 & 3 $\pm$ 21 & 0.74 $\pm$ 0.05 & 3.0 $\pm$ 1.8 \\
60361.11500 & 0.40 $\pm$ 0.20 & 21 $\pm$ 14 & 0.24 $\pm$ 0.07 & 33.5 $\pm$ 8.5 & 0.44 $\pm$ 0.05 & 33.3 $\pm$ 3.3 \\
\multicolumn{7}{c}{DHS} \\
60390.23863 & 0.13 $\pm$ 0.07 & 13 $\pm$ 16 & 0.48 $\pm$ 0.09 & 67.3 $\pm$ 5.1 & 0.09 $\pm$ 0.08 & $-20$ $\pm$ 23 \\  
\hline
\end{tabular}
\tablefoot{
The IS polarization estimate used stars 2, 4, and 5.   
Observations during the SS have been averaged over two days and during the DHS have been averaged over three days before computation of the intrinsic polarization. 
For non-significant detection of polarization below $1\sigma$, the $3\sigma$ upper limits on the PD are given. 
For the significance of polarization detection exceeding $1\sigma$, the PAs are given with the formal error from Eq.~\eqref{eq:errPA}. }
\end{table*}

\end{appendix}

\end{document}